\documentclass[fdp,a4paper,fleqn%
]{w-art}
\usepackage{times,cite,w-thm}
\usepackage{amsmath}
\usepackage{amssymb}
\theoremstyle{plain}

\theoremstyle{definition}

\usepackage[]{graphicx}
\usepackage{psfrag}
\begin{document}
\bibliographystyle{fdp}
\DOIsuffix{theDOIsuffix}
\Volume{55}
\Month{01}
\Year{2007}
\pagespan{1}{}
\Receiveddate{XXXX}
\Reviseddate{XXXX}
\Accepteddate{XXXX}
\Dateposted{XXXX}
{\bf Preprint}: LA-UR-12-20346


\title[Non-Hermitian approach \dots]{Non-Hermitian approach for modeling of noise-assisted quantum electron transfer in photosynthetic complexes}


\author[Alexander I.~Nesterov]{Alexander I.~Nesterov\inst{1,}%
  \footnote{Corresponding author\quad E-mail:~\textsf{nesterov@cencar.udg.mx},
           }}
\address[\inst{1}]{Departamento de F{\'\i}sica, CUCEI, Universidad de Guadalajara, Av. Revoluci\'on 1500, Guadalajara, CP 44420, Jalisco, M\'exico}
\author[Gennady P.~Berman]{Gennady P.~ Berman\inst{2,} \footnote{E-mail:~\textsf{gpb@lanl.gov}}}
\address[\inst{2}]{Theoretical Division, MS-B213, Los Alamos National Laboratory, Los Alamos, NM 87544, USA}
\author[Alan R.~Bishop]{Alan R.~Bishop\inst{3,}\footnote{E-mail:~\textsf{ arb@lanl.gov}}}
\address[\inst{3}]{STE, MS-A127, Los Alamos National Laboratory, Los Alamos,  NM, 87544, USA}

\begin{abstract}
We model the quantum electron transfer (ET) in the photosynthetic reaction center (RC), using a non-Hermitian Hamiltonian approach. Our model includes (i) two protein cofactors, donor and acceptor, with discrete energy levels and (ii)  a third protein pigment (sink) which has a continuous energy spectrum. Interactions are introduced between the donor and acceptor, and between the acceptor and the sink, with noise acting between the donor and acceptor. The noise is considered classically (as an external random force), and it is described by an ensemble of two-level systems (random fluctuators). Each fluctuator has two independent parameters, an amplitude and a switching rate. We represent the noise by a set of fluctuators with fitting parameters (boundaries of switching rates), which allows us to build a desired spectral density of noise in a wide range of frequencies.  We analyze the quantum dynamics and the efficiency of the ET as a function of (i) the energy gap between the donor and acceptor, (ii) the strength of the interaction with the continuum, and (iii) noise parameters.  As an example, numerical results are presented for the ET through the active pathway in a quinone-type photosystem II RC.

\end{abstract}
\maketitle                   






\section{Introduction}

Nature has evolved photosynthetic organisms to be extremely complex bio-engines that capture visible light in their peripheral light-harvesting complexes (LHCs) and transfer excited-state energy (as excitons) through the proximal LHC of  photosystem II (PSII) and photosystem I (PSI) to the RCs. The primary charge separation occurs in the RC (which works as a battery), leading to the formation of  an electrochemical gradient \cite{BER1,GSR,XSG,Psr}. During the past two decades,  crystallographic structures for many photosynthetic complexes (PCs), including the LHCs and RCs, have been determined to a resolution of 2.5-3 {\AA} \cite{KHM,HDR,CWW}. (See also references therein.)

 Like all engines, PCs operate in a thermal environment at ambient temperature and in the presence of external ``classical" sources of noise \cite{MRSN,XSK,SWB,MFL,BPM,DBJ,NB1b}. In spite of this, recent experiments based on two-dimensional laser-pulse femtosecond photon echo spectroscopy revealed a long-lived exciton-electron quantum coherence in PCs such as the Fenna-Matthews-Oslov (FMO) and marine algae \cite{ECR,CWWC,PHFC}. Mainly, this occurs because the dynamics of the ET  is so rapid (some picoseconds) that the thermal fluctuations and external noise are unable to significantly destroy quantum coherence. Consequently, the exciton/electron dynamics in LHCs-RCs must be described using quantum-mechanical methods \cite{IFG,RMKL,CFMB,PPM,AMS}. (See also references therein.) An important consequence of this is the high ET efficiency of the peripheral antennae complexes (close to 100 \%).

 As an example, Fleming and colleagues \cite{IFG} have modeled quantum coherence effects in the bacterial FMO LHC  by (i) using a tight-binding model (TBM) for exciton dynamics and (ii) introducing an empirical thermal relaxation function having an exponential form, in order to describe the high efficiency of exciton energy transport. Usually, in the TBM the exciton/electron ET dynamics in LHCs-RCs is described in the single exciton/electron approximation (due to limited sunlight intensities), with   $N$ ( $N=7$ for the FMO in \cite{IFG}) being the total number of discrete pigments/sites. (Note that  more complicated models which account for exciton and charged states can also be used \cite{AMS}.) In this case, each pigment, $n \,(n=1,\dots,N)$,  is represented by a two-level system with states $|0_n\rangle$ (unoccupied) and $|1_n\rangle$ (occupied). The total Hamiltonian is $H_{tot} = H_e + H_{ph} + H_{el-ph} $ \cite{IFG}. The first term is the Hamiltonian of exciton/electron states of the pigments in the site representation: $H_e= \sum_1^N E_n |n\rangle \langle n| + \sum^N_{m\neq n}  V_{mn}|m\rangle\langle  n |  $, where $E_n$ is the site energy, and $V_{mn}$  denotes the coupling between the  $n$-th and  $m$-th pigments. The term $H_{ph}$  describes the thermal phonons provided by the protein environment, and the third term describes the interaction between pigments and the thermal phonons. It was numerically demonstrated  in \cite{IFG}, that in the FMO complex, quantum coherent ET is an adequate way to describe the energy transport dynamics.

Usually, there are two different approaches which are used to describe the influence of the protein environment on the ET. One is based on  the thermal environment \cite{XSK}. In this case, the environment acts self-consistently on the electron system and, in combination with the transition amplitudes between sites/pigments, provides the ET rates between the sites and the Gibbs equilibrium state for the LHC-RC subsystem. The other approach is based on considering an external ``classical" noise \cite{PPM} provided by the protein vibrations. This approach results in a transfer rate for the electron, but does not lead to Gibbs equilibrium states. The choice of  approach depends on the concrete experimental situation which the theoretical model is intended to describe.

 In this paper we use the second approach, modeling the noise by an ensemble of fluctuators \cite{NB1b}. To simplify our description, we introduce a set of fluctuators with fitting parameters (boundaries of switching rates between relatively slow and fast fluctuators), which allows us to build a desired spectral density of noise in a wide range of frequencies.  In particular, the spectral density of noise, used in this paper, includes the components of white noise,  $1/f$ noise, and high-frequency noise. We demonstrated in \cite{NB1b} that this approach successfully described the experiments \cite{YHNN} on the quantum dynamics of superconducting qubits. Here we consider the simplest model of ET in a quinone-type  active pathway of the PSII RC. Our model includes  two protein cofactors (donor and acceptor) with discrete energy levels, with the acceptor being embedded in a third protein pigment (sink) that has a continuous energy spectrum. In \cite{CFMB} an additional sink reservoir was empirically introduced in order to describe the high ET efficiency in the FMO complex. A sink reservoir was also introduced phenomenologically in \cite{PPM} to describe the asymmetry of two branches of the ET in the photosynthetic PSII RC, and in \cite{THA} to describe the dynamics of excitons in photosynthetic systems. In our model, the influence of the sink is described self-consistently, using a non-Hermitian Hamiltonian approach. We include the interactions between the donor and acceptor, and between the acceptor and the sink. The classical noise acts only between the donor and acceptor. We analyze the dynamics and the efficiency of the ET as a function of the energy gap between donor and acceptor, the strength of interaction with continuum, and the noise parameters.  We calculate explicitly the ET rate and efficiency as a function of parameters. We demonstrate the regimes in which noise assists the ET efficiency (in particular, in which the influence of  noise  significantly increases the efficiency of the ET from the ``donor-acceptor" subsystem to the sink).

 Our paper is organized as follows. In Section II, using the Feshbach projection method, we introduce an effective non-Hermitian Hamiltonian to describe the RC consisting of the donor and acceptor coupled to the sink. In Section III, we study the dynamics of the electron transfer without noise. In Section IV, we study the decoherence effects caused by the  classical noise on the ET efficiency. In the Section V, we discuss the obtained results. In the Appendices some important formulae are presented.

\section{Model description}

\begin{figure}[tbh]
\begin{center}
\scalebox{0.325}{\includegraphics{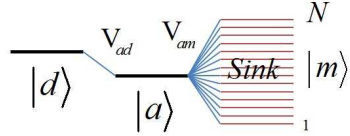}}
\end{center}
\caption{A reaction center consisting of the donor and acceptor discrete energy levels, with the acceptor coupled to a sink reservoir with a continuous spectrum.
\label{S1a}}
\end{figure}

We consider a model (``building block" of the LHCs-RCs) of the RC with  three sites (protein pigments):  the first site, $|d\rangle$, is the electron donor (with the energy $E_d$ ), the second site, $|a\rangle$, is the electron acceptor (with the energy $E_a$ ), and the third site is a ``sink", with a continuous spectrum. We assume that the acceptor is coupled to the sink, which we first model by a large number of discrete and nearly degenerate energy levels, $N \gg 1$ (Fig. \ref{S1a}).
The Hamiltonian of this system can be written as,
\begin{align}
H_t = E_d|d\rangle \langle d|+
E_a|a\rangle \langle a| + \frac{V}{2}(|d\rangle \langle a|+ |a\rangle \langle d|) \nonumber \\
+\sum^N_{n=1}E_n |n\rangle\langle  n | + \sum^N_{m=1} \big( V_{am}|a\rangle\langle  m | + V_{ma}|m\rangle\langle a |\big).
\end{align}

The total Hilbert space can be divided into two orthogonal subspaces generated by two projection operators, $P= |d\rangle \langle d|+ |a\rangle \langle a|$ and $Q= \sum_1^N(|n\rangle \langle n|)$, where the $P$-space is associated with the donor-acceptor levels and the $Q$-space is associated with the sink. These projection operators have the following properties: $P+Q=1$, $P^{2} =P$, $Q^{2} =Q$ and $PQ=QP=0$. Then, using the Feshbach projection method \cite{RI,RI2,RI3,VZ}, we obtain the effective non-Hermitian Hamiltonian that describes only the ``donor-acceptor" subsystem,
\begin{align}\label{H1c}
 \tilde{\mathcal H}= E_d|d\rangle \langle d|+ (E_a+ \Delta (E) - \frac{i}{2}\Gamma_a(E))|a\rangle \langle a|+
\frac{V}{2}(|d\rangle \langle a|+ |a\rangle \langle d|),
\end{align}
where
\begin{align}\label{Delta}
\Delta(E) -  \frac{i}{2}\Gamma_a(E) =  \sum_n \frac{|V_{an}|^2}{E - E_n + i\delta}.
\end{align}

To proceed further, we assume that the sink is sufficiently dense, so that one can perform an integration instead of a summation. Then we have,
\begin{align}
\Delta(E) -  \frac{i}{2}\Gamma_a(E) =  \int \frac{|V_{an}|^2 g(E_n)dE_n}{E - E_n + i\delta},
\end{align}
where $g(E_n)$ is the density of states of the sink. One can show that \cite{SM}
\begin{align}
\Delta(E) = {\mathcal P} \int \frac{|V_{an}|^2g(E_n)dE_n}{E - E_n}, \\
\Gamma_a(E) =  2\pi\int |V_{an}|^2g(E_n)\delta(E - E_n) dE_n,
\end{align}
where $\mathcal P$ denotes the principal value of the integral.

The exact dynamical evolution of the whole quantum system (RC) is
described by the Schr\"odinger equation (we set $\hbar =1$),
\begin{align}\label{S}
i\frac{\partial \psi (t)}{\partial t}= H_t\psi (t).
\end{align}
We assume that at $t=0$ the system is populated in the $P$-space. If the Q-space represents a smooth continuum (which is assumed below) one can neglect the dependence of $\Delta(E)$  and $\Gamma_a(E)$ on $E$. Denoting these functions as $\Delta$ and $\Gamma_a$,  one can find that the dynamics of the donor-acceptor (intrinsic) states can be described by the following Schr\"odinger equation with the effective non-Hermitian Hamiltonian, $\tilde{\mathcal H}$:
\begin{align}\label{S1b}
   i\frac{\partial \psi _{p} (t)}{\partial t}= \tilde{\mathcal H} \psi _{p} (t),
\end{align}
where $\psi _{p} (t)=P\psi (t)$. Further, it is convenient to rewrite $\tilde{\mathcal H}$ as $ \tilde{\mathcal H}= {\mathcal H}- i \mathcal W$, where
\begin{align}
{\mathcal H} = \varepsilon_d|d\rangle \langle d|+ \varepsilon_a|a\rangle \langle a|+ \frac{V}{2}(|d\rangle \langle a|+ |a\rangle \langle d|)
\end{align}
is the dressed donor-acceptor Hamiltonian, $\mathcal W = ({1}/{2})\Gamma_a|a\rangle \langle a|$, with $\varepsilon_d=E_d$ and $ \varepsilon_a= E_a+ \Delta  $.

We define $\rho_t(t)$ to be the density matrix that satisfies the conventional equation of motion with the total Hamiltonian, $H_t$: $i\dot{ \rho}_t = [H_t,\rho]$.
Next, we introduce the projected density matrix as $\rho(t) = P\rho_t(t) P$. Then, one can show that $\rho(t)$ satisfies the Liouville  equation,
\begin{eqnarray}\label{DM1}
   i \dot{ \rho} = [\mathcal H,\rho] - i\{\mathcal W,\rho\},
\end{eqnarray}
where $\{\mathcal W,\rho\}= \mathcal W\rho +\rho\mathcal  W$.

Assume now that the quantum system under consideration interacts with the environment. We use the reduced density matrix approach to describe this interaction. To include into the description of the system  both processes of decoherence and tunneling to the continuum, we introduce the following generalized master equation, $ i\dot{ \rho} = [\mathcal H,\rho] + {\mathcal L}\rho-  i\{\mathcal W,\rho\}$, where $\mathcal H$ is the dressed Hamiltonian, and  the Lindblad operator, $\mathcal L$, describes the coupling to the environment. The commutator of the density operator, $\rho$, with the Hamiltonian, $\mathcal H$, is the coherent part of evolution, and the remaining part corresponds to the decoherence process causes by the interaction with the environment.

\section{Tunneling to the sink}

We consider here the quantum dynamics of the ET from the donor $|d\rangle$ ($|1\rangle$) to the acceptor $|a\rangle$ ($|2\rangle$) coupled to the sink. We assume that the acceptor is coupled to the  $N$-level sink reservoir and that the corresponding Hilbert subspace is dense and smooth. For description of the tunneling from the acceptor to the sink we use the Feshbach projection method described above. This yields the following effective non-Hermitian Hamiltonian:
\begin{align}\label{H3}
\tilde{ {\mathcal H}}= \frac{\tilde\lambda_0}{2} \left(\begin{array}{cc} 1& 0\\ 0& 1 \end{array}\right) + \frac{1}{2} \left(\begin{array}{cc} {\varepsilon + i\Gamma}&V\\V& {-\varepsilon -i\Gamma} \end{array}\right),
\end{align}
where $\tilde\lambda_0= \varepsilon_{1} +\varepsilon_{2} -i \Gamma$, $\varepsilon = \varepsilon_{1} -\varepsilon_{2}$ ($\varepsilon_n$ is the renormalized energy), $\Gamma = \Gamma_a/2 $, with $\Gamma_a$ being the relaxation rate from the acceptor to the sink.

{\em Region of parameters.} The model involves various parameters, which are only partially known. For concreteness of the numerical simulations, our choice of the parameters is based on the data taken  for the ET through the active pathway in the quinone-type of the photosystem II RC \cite{LVD} (in the units $\hbar =1$): $\varepsilon = 60 \rm ps^{-1}$ and $ 10 \rm ps^{-1} < V < 40 \rm ps^{-1}$. The parameter $\Gamma$ is varied in the interval: $ 1 \rm ps^{-1}< \Gamma < 5 \rm ps^{-1}$. But also other values of parameters, $\varepsilon$ and $V$, are used in our numerical simulations.  (Note, that the values of parameters in energy units can be obtained by multiplying our values by $\hbar\approx 6.58\times 10^{-13}\rm meVs$. For example, $\varepsilon = 60\; \rm ps^{-1}\approx 40\rm meV$.)

In what follows we assume that initially the quantum system occupies the upper level (donor), $\rho_{11}(0)=1$ ($\rho_{22}(0)=0$). Then, for the diagonal component of the density matrix the solution of the Liouville equation (\ref{DM1}) is given by (for details see Appendix A),
\begin{align}
\rho_{11}(t)  = {e^{-\Gamma t}} \bigg |\Big(\cos\frac{\Omega t}{2} - i\cos\theta\sin\frac{\Omega t}{2}\Big)\bigg|^2, \quad
\rho_{22}(t)  = {e^{-\Gamma t}} \bigg |\sin\theta\sin\frac{\Omega t}{2}\bigg |^2,
\label{P3d}
\end{align}
where $\Omega= \sqrt{V^2 +(\varepsilon + i\Gamma )^2}$ is the complex Rabi frequency, $\cos\theta = (\varepsilon + i\Gamma )/\Omega$, and $\sin\theta = V/\Omega$.
\begin{figure}[tbh]
\begin{center}
\scalebox{0.325}{\includegraphics{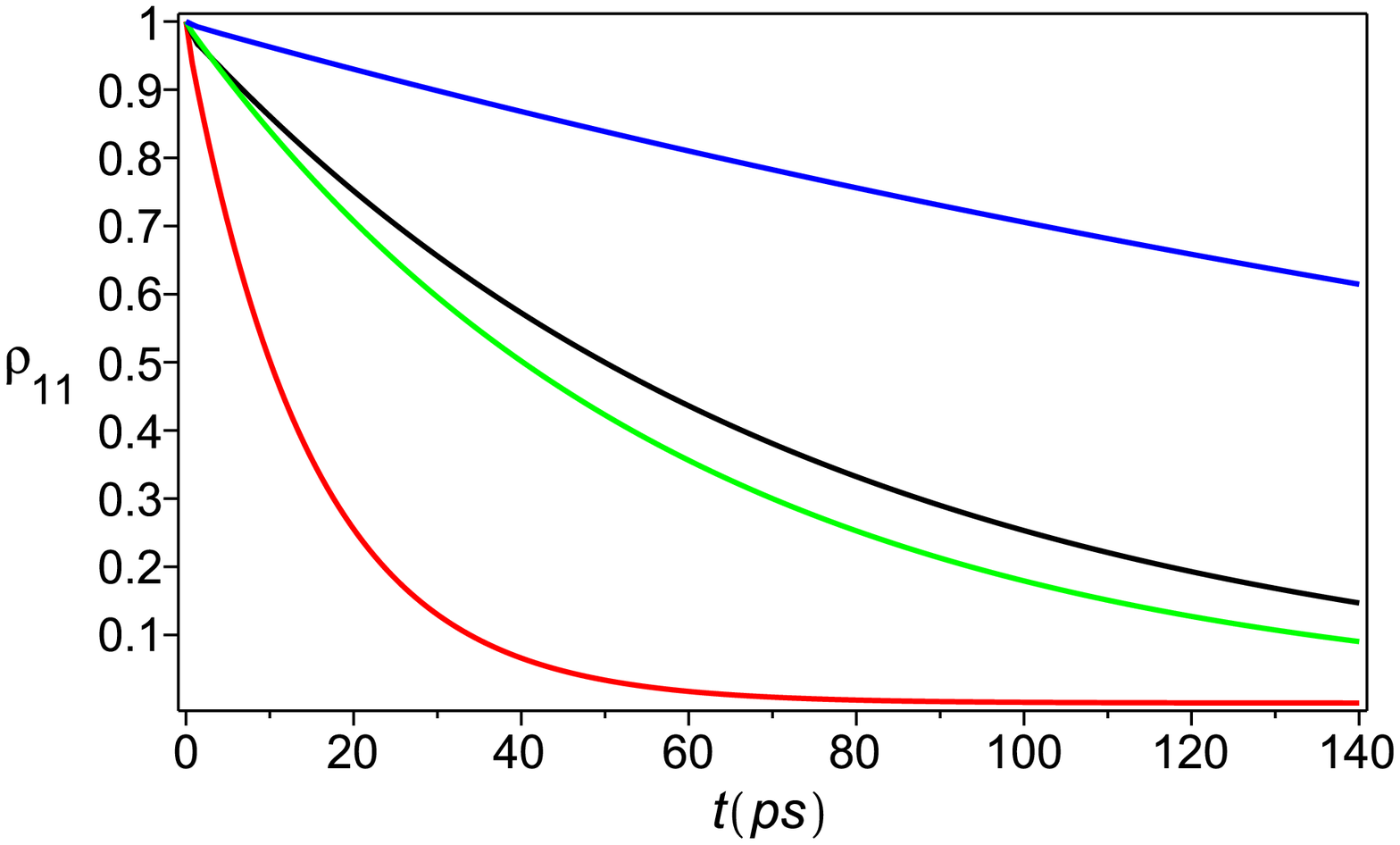}}
\scalebox{0.3}{\includegraphics{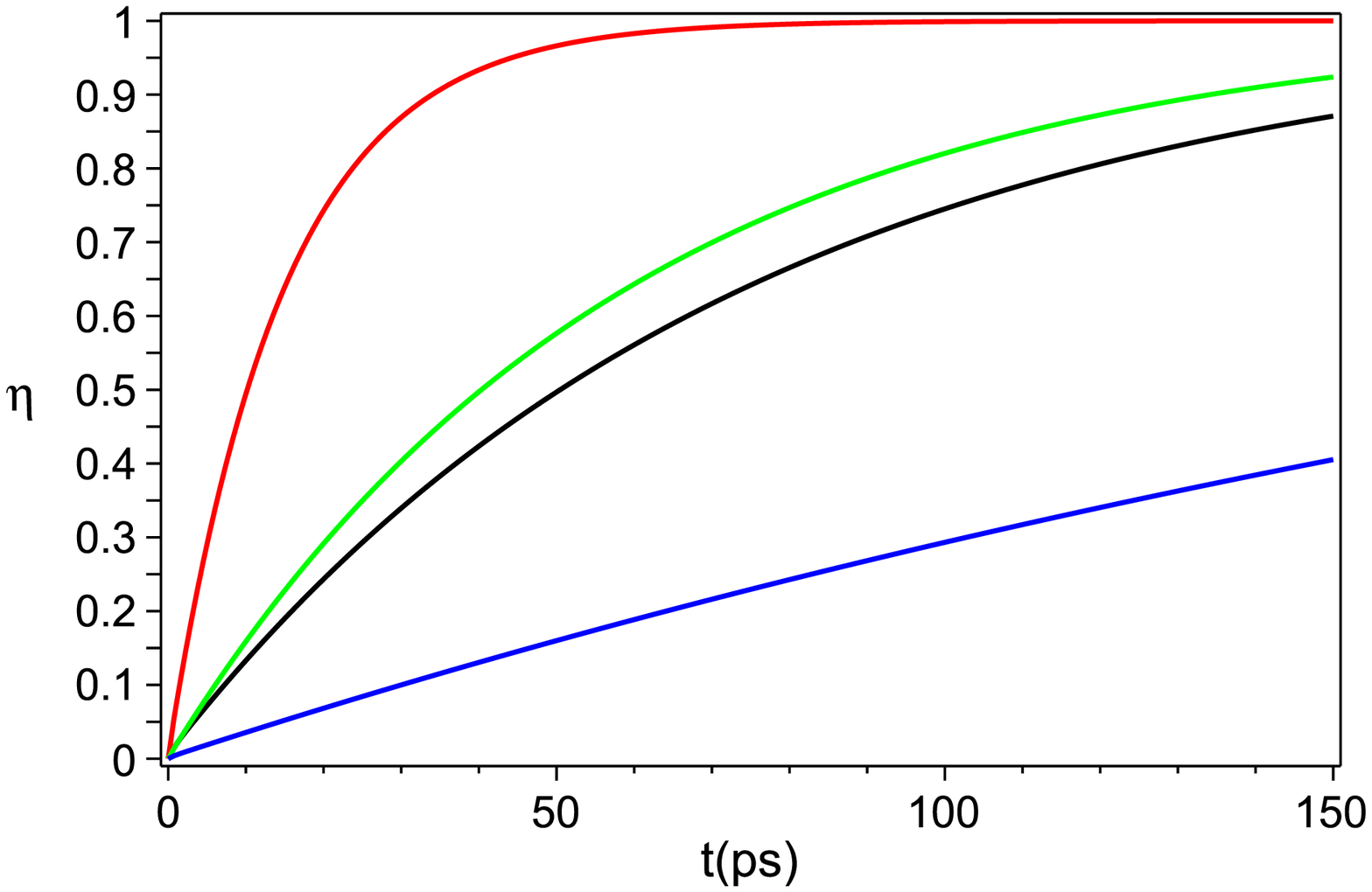}}
\end{center}
\caption{ Left panel: The time dependence of the population of the donor site.
Right panel: ET efficiency. The parameters are the following: blue line ($\Gamma=1\; \rm ps^{-1}$, $  V = 10 \rm ps^{-1} $), black line ($\Gamma=1 \rm ps^{-1}$, $  V = 20 \rm ps^{-1} $), green line ($\Gamma=5 \rm ps^{-1}$, $  V = 10 \rm ps^{-1} $), red line ($\Gamma=5 \rm ps^{-1}$, $  V = 20 \rm ps^{-1} $). In all cases $\varepsilon = 60 \rm ps^{-1} $.
\label{R1}}
\end{figure}

The ET efficiency can be defined as the integrated probability of trapping the electron in the sink \cite{RMKL,CDCH},
\begin{eqnarray}
\eta(t) = 2\Gamma \int_0^t \rho_{22}(\tau)d \tau.
\label{Eq16}
\end{eqnarray}
Setting $\Omega= \Omega_1 +i \Omega_2$ and performing the integration, we obtain for the ET efficiency,
\begin{align}\label{Eq17}
 \eta(t) = 1- \frac{ e^{-\Gamma t}}{\Gamma(\Omega^2_1 +\Omega^2_2)}\big((\Gamma^2 + \Omega_1^2) (\Gamma\cosh{\Omega_2 t} + \Omega_2\sinh{\Omega_2 t}) \nonumber \\
 - (\Gamma^2 - \Omega_2^2) (\Gamma\cos{\Omega_1 t} -\Omega_1\sin{\Omega_1 t} )\big).
\end{align}
This yields the following large-time asymptotic behavior:
\begin{align}\label{Eq17}
 \eta(t) \sim 1- \frac{(\Gamma - \Omega_2 )(\Gamma^2 + \Omega_1^2)}{2\Gamma(\Omega^2_1 +\Omega^2_2)} e^{-(\Gamma +\Omega_2)t} .
\end{align}
The numerical results are presented in Fig. \ref{R1}. As one can see, for these values of parameters, and without the action of noise, the ET efficiency approaches a value close to 1 for relatively large times, $t>150$ ps.

Let us consider now the flat redox potential, $\varepsilon =0$. From the relation $\Omega_1\Omega_2 = \varepsilon \Gamma$, it follows that for $\varepsilon = 0$ there are two possibilities: (i) $\Omega_1=0$, $\Omega_2 = \sqrt{\Gamma^2 - V^2}$ ($V < \Gamma)$; and (ii) $\Omega_2 = 0$, $\Omega_1 = \sqrt{ V^2 - \Gamma^2}$  ($V > \Gamma)$. Using these results we obtain,
\begin{align}\label{Eq17a}
 \eta(t) =\left \{ \begin{array}{ll}
 1- \displaystyle \frac{ e^{-\Gamma t}}{\Omega^2_1 }\big((\Gamma^2(1- \cos{\Omega_1 t}) + \Omega_1 ( \Omega_1- \Gamma \sin{\Omega_1 t} )\big), &  V > \Gamma  \\
  1- \bigg(1+ \Gamma t + \bigg(\displaystyle \frac{\Gamma t}{2}\bigg)^2\bigg) e^{-\Gamma t}, & V = \Gamma \\
   1- \displaystyle\frac{ e^{-\Gamma t}}{\Omega^2_2 }\big((\Gamma^2(\cosh{\Omega_2t}-1) + \Omega_2 ( \Omega_2 +\Gamma \sinh{\Omega_1 t} )\big), & V < \Gamma
                \end{array}
                \right.
 \end{align}
This yields the following asymptotic behavior for the ET efficiency, $\eta(t)$ ($\Gamma t \gg 1$):
\begin{align}\label{Eq17d}
\eta(t) \sim\left \{ \begin{array}{ll}
 1- \displaystyle \frac{\Gamma^2}{2\Omega_1^2 } e^{-\Gamma t}, &  V > \Gamma  \\
  1- \bigg(\displaystyle \frac{\Gamma t}{2}\bigg)^2 e^{-\Gamma t}, & V = \Gamma \\
  1- \displaystyle \frac{\Gamma^2 }{2\Omega_2^2 }e^{-(\Gamma - \Omega_2 )t} , & V < \Gamma
  \end{array}
  \right.
\end{align}
Comparing Eqs. (\ref{Eq17}) - (\ref{Eq17d}), we conclude that the highest ET efficiency is obtained for the flat redox potential ($\varepsilon =0$), and $V > \Gamma$.

\subsection{Quantum evolution in the vicinity of the exceptional point}

For the Hermitian Hamiltonian, the coalescence of eigenvalues
results in different eigenvectors and the related degeneracy,
referred to as a ``conical intersection", is known also as
a ``diabolic point" \cite{B0}. However, in a quantum mechanical system
governed by a non-Hermitian Hamiltonian merging not
only of eigenvalues of the Hamiltonian but also of the associated
eigenvectors can occur. In this case, the point of coalescence  is called
an ``exceptional point" (EP). At the EP, the eigenvectors merge,
forming a Jordan block. (For a review and references, see,
e.g., \cite{B}.)

\begin{figure}[tbh]
\begin{center}
\scalebox{0.325}{\includegraphics{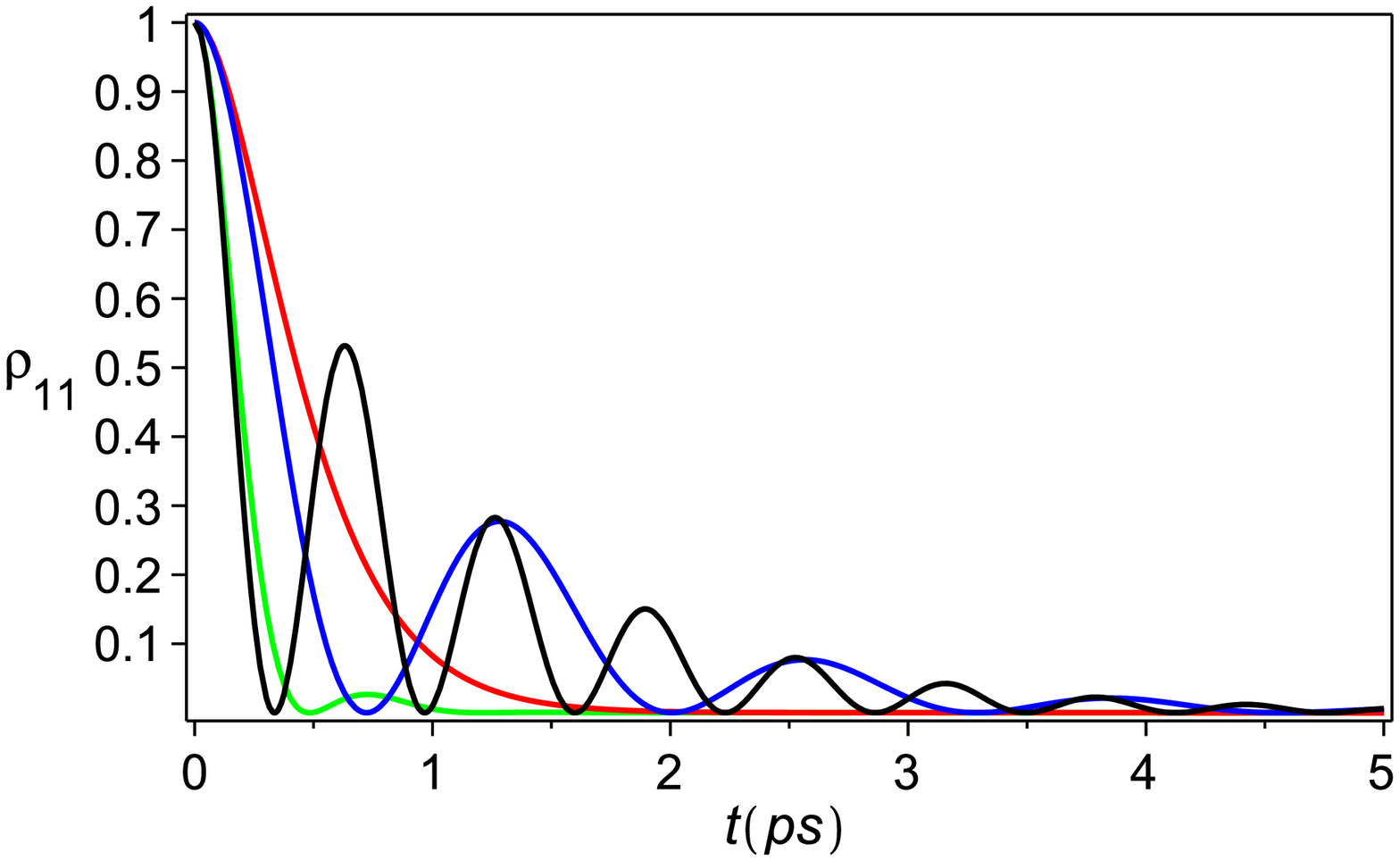}}
\scalebox{0.33}{\includegraphics{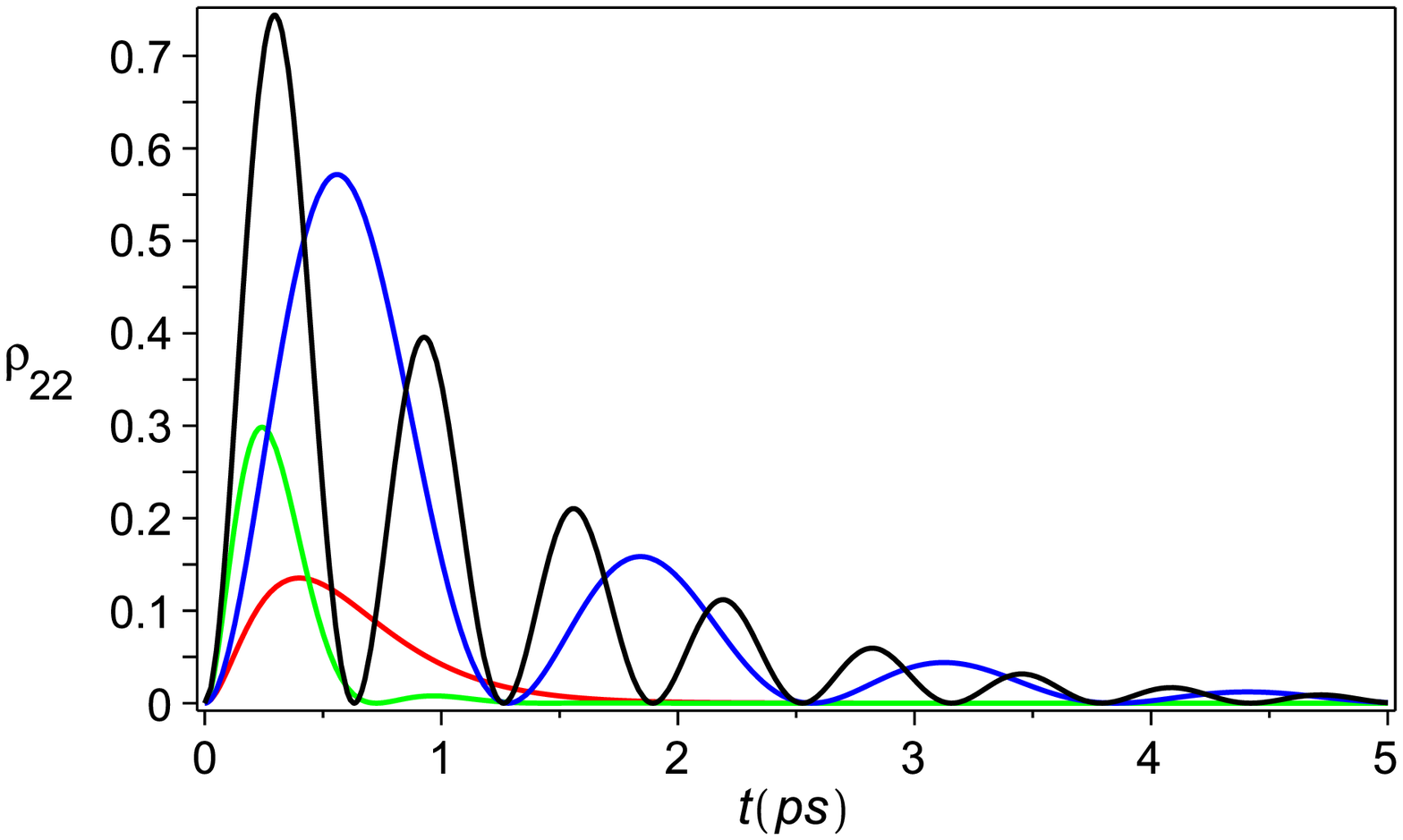}}
\end{center}
\caption{Time dependence of site populations in the vicinity of the EP (red line) for the flat redox potential ($\varepsilon =0$). The parameters are chosen as the following: blue line ($\Gamma=1\; \rm ps^{-1}$, $  V = 5 \rm ps^{-1} $), black line ($\Gamma=1 \rm ps^{-1}$, $  V = 10 \rm ps^{-1} $), green line ($\Gamma=5 \rm ps^{-1}$, $  V = 10 \rm ps^{-1} $), red line ($\Gamma=5 \rm ps^{-1}$, $  V = 5 \rm ps^{-1} $) corresponds to the exceptional point.
\label{Ep1}}
\end{figure}

In the effective two-level system under consideration, the EP is defined by equation $\Omega =0$. This yields $\varepsilon =0$ and $V^2 - \Gamma^2 =0$. To study tunneling to the sink near a degeneracy, we assume the flat dressed redox potential, $\varepsilon = 0$. Then, there are two different regimes of the ET depending on the relative values of $V$ and $\Gamma$. For $V > \Gamma$, we have a {\em coherent} tunneling process (with oscillating probabilities, see Fig.~\ref{Ep1}),
\begin{figure}[tbh]
\begin{center}
\scalebox{0.35}{\includegraphics{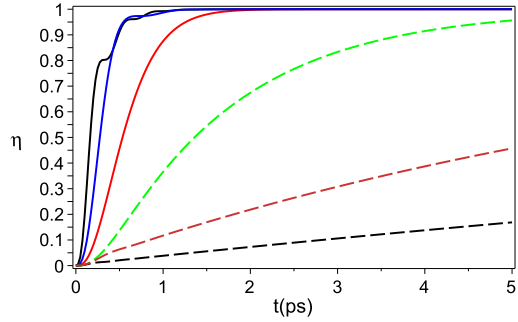}}
\end{center}
\caption{The ET efficiency in the vicinity of the EP ($  \Gamma = 5 \rm ps^{-1} $). Black line ($ V = 20 \rm ps^{-1} $, $  \varepsilon = 0 \rm ps^{-1} $), blue line ($ V = 10 \rm ps^{-1} $, $  \varepsilon = 0 $). Red line: tunneling at the EP ( $ V = 5 \rm ps^{-1} $, $  \varepsilon = 0 $). Black dashed line ( $ V = 2.5 \rm ps^{-1} $, $  \varepsilon =  20 \rm ps^{-1} $). Green dashed line ( $ V = 2.5 \rm ps^{-1} $, $  \varepsilon =  0 $). Orange dashed line ($ V = 2.5 \rm ps^{-1} $, $  \varepsilon =  10 \rm ps^{-1}  $).
\label{Ep1a}}
\end{figure}

\begin{eqnarray}\label{T4}
\rho_{11} =  e^{-\Gamma t}\Big(\cos\frac{\Omega_0 t}{2}+ \frac{\Gamma}{\Omega_0}\sin\frac{\Omega_0 t}{2}\Big)^2 , \quad
\rho_{22}=   e^{-\Gamma t} \frac{V^2}{\Omega_0^2}\sin^2\frac{\Omega_0 t}{2},
\end{eqnarray}
where $\Omega_0= |V^2 - \Gamma^2|^{1/2}$ denotes the Rabi frequency.

On the other hand, for $V < \Gamma$, the tunneling becomes {\em incoherent}, without probability oscillations,
\begin{eqnarray}\label{T5}
\rho_{11} =  e^{-\Gamma t}\Big(\cosh\frac{\Omega_0 t}{2}+ \frac{\Gamma}{\Omega_0}\sinh\frac{\Omega_0 t}{2}\Big)^2 , \quad
\rho_{22}=   e^{-\Gamma t} \frac{V^2}{\Omega_0^2}\sinh^2\frac{\Omega_0 t}{2}.
\end{eqnarray}

At the EP, $\Omega_0 = 0$, and both regimes coincide. (See Fig.~\ref{Ep1}, red curve.) In this case, we have the following solutions for the probabilities,
\begin{eqnarray}\label{EP2}
 \rho_{11}(t)  = {e^{-\Gamma t}} \bigg(1 +\frac{\Gamma t}{2}\bigg)^2, \quad
 \rho_{22}(t)  = {e^{-\Gamma t}} \bigg (\frac{\Gamma t}{2}\bigg )^2.
\end{eqnarray}

The results of numerical simulations of the ET efficiency in the vicinity of the EP are shown in Fig.~\ref{Ep1a}. One can see that, for the chosen parameters, the ET efficiency can approach a value close to 1 for short times, $\sim 2$ ps. Note that the coherent tunneling regime ($V>\Gamma$) is more effective for approaching a high ET efficiency for short times. (See Fig.~\ref{Ep1a}, black and blue curves.)

\section{Noise-assisted electron transfer}

In this section, we consider ET  from the donor, $|1\rangle$, to the acceptor, $|2\rangle$, coupled to the sink, in the presence of classical noise. Then, the effective non-Hermitian Hamiltonian (\ref{H3}) takes the form
\begin{eqnarray} \label{Eq2b}
 \tilde{\mathcal H}= \sum_n \varepsilon_n |n\rangle\langle  n |+ \sum_{m,n} \lambda_{mn}(t))|m\rangle\langle  n | + \frac{V}{2}\sum_{m \neq n}  |m\rangle\langle  n | - i\Gamma |2\rangle \langle 2|, \quad m,n = 1,2,
\end{eqnarray}
where $\lambda_{mn}(t))$ describes the noise. In our approach, we use a spin-fluctuator model of noise with the number of fluctuators, ${{\mathcal N}} \gg 1$ \cite{GABS,BGA,NB1b}. The diagonal matrix elements of noise, $\lambda_{nn}$, are responsible for decoherence, and the off-diagonal matrix elements, $\lambda_{mn}$ ($ m \neq n$), lead to the relaxation processes.

The approximate equations of motion for the  average diagonal components of the density matrix are given by (for details see Appendix B)
\begin{align} \label{N4}
\frac{d}{dt}{\langle{\rho}}_{11}(t)\rangle =-{\mathfrak R}(t)\big(\big\langle{\rho}_{11}(t)\big\rangle -\big\langle{\rho}_{22}(t)\big\rangle\big) + {\mathcal O}(|V|^4), \\
\frac{d}{dt}{\langle{\rho}}_{22}(t)\rangle ={\mathfrak R}(t)\big(\big\langle{\rho}_{11}(t)\big\rangle -\big\langle{\rho}_{22}(t)\big\rangle\big) - 2\Gamma \langle {\tilde \rho}_{22}(t)\rangle + {\mathcal O}(|V|^4),
\label{N5}
\end{align}
where the average $\langle\; \rangle$ is taken over the random process describing noise, and
\begin{align}\label{N6}
{\mathfrak R}(t) =\frac{1}{4}\int_0^t e^{-\Gamma (t- t')}\big(\big\langle{\tilde V}(t){\tilde V}(t') \big\rangle + \big\langle{\tilde V}(t'){\tilde V}(t)\big\rangle\big)dt'.
\end{align}

{\em The model of noise.} In the following, we restrict ourselves to consider only diagonal noise effects, assuming that the noisy environment is the same for the donor and acceptor sites (collective noise). Then, one can write $\lambda_1(t) = g_1 \xi(t) $ and $\lambda_2(t) = g_2 \xi(t) $, where $\xi(t)$ is a random variable describing the stationary noise with the correlation function, $\chi(t-t')=\langle \xi(t)\xi(t')\rangle$, and $g_{1,2}$ are the interaction constants. We describe the noise by a spin-fluctuator model with the number of fluctuators, ${\mathcal N} \gg 1$,  with the correlation function, $\chi(\tau)$,  given by \cite{NB1b} \begin{eqnarray}
\chi(\tau) = \sigma^2 A\Big(E_1(2\gamma_m \tau) - E_1(2\gamma_c \tau) \Big ), \quad \tau = |t-t'|,
\end{eqnarray}
where $E_n(z)$ denotes the Exponential integral \cite{abr}, $A= 1/\ln(\gamma_c/\gamma_m)$ and $\chi(0)= \sigma^2$. The spectral density of the noise, defined as
\begin{eqnarray}
S(\omega) = \frac{1}{\pi}\int\limits_{0}^{\infty}\chi(\tau) \cos(\omega \tau) d\tau,
\label{Sf1c}
\end{eqnarray}
is given in \cite{NB1b},
\begin{align}\label{SF1}
S(\omega) = \frac{\sigma^2 }{\pi\omega \ln(\gamma_c/\gamma_m)}\bigg( \arctan \Big(\frac{\omega}{2\gamma_m}\Big)- \arctan\Big(\frac{\omega}{2\gamma_c}\Big)\bigg).
\end{align}
This yields the following asymptotic behavior of $S(\omega)$:
\begin{eqnarray}
S(\omega) \approx \left \{\begin{array}{ll}
\displaystyle\frac{\sigma^2 }{2\pi \gamma_m \ln(\gamma_c/\gamma_m) }\bigg(1- \frac{\gamma_m}{\gamma_c}\bigg),&   \omega \ll2\gamma_m ,\\
\\
\displaystyle\frac{\sigma^2 }{2  \omega\ln(\gamma_c/\gamma_m)},& 2\gamma_m \ll  \omega \ll 2\gamma_c, \\
\\
\displaystyle\frac{2\sigma^2 \gamma_c (1-\gamma_m/\gamma_c)}{\pi \omega^2 \ln(\gamma_c/\gamma_m)},&   \omega \gg 2\gamma_c,
\label{Sf1d}
\end{array}
\right.
\end{eqnarray}
 where $\gamma_m$ and $\gamma_c$ ($\gamma_m\ll\gamma_c$) indicate the boundaries of the switching rates in the ensemble of random fluctuators. As one can see from Eqs. (\ref{Sf1d}), for  $\omega \ll 2\gamma_m $ the spectral density of noise, $S(\omega)$, describes the white noise. In the interval of frequencies, $2\gamma_m \ll  \omega \ll 2\gamma_c$, one has the $1/f$ noise: $S\sim 1/f$ ($f=\omega/2\pi$). And for   $ \omega \gg 2\gamma_c$, we obtain the Lorentzian spectrum.

{\em Choice of parameters.} The correlation function includes, besides the amplitude, $\sigma$,  two fitting parameters: $\gamma_m$ and $\gamma_c$.  Taking into account available theoretical and experimental data \cite{TMT,CBC,JBS}, we have chosen in our numerical simulations the following parameters: $ 2\gamma_m = 10^{-4}\rm ps^{-1} $, $ 2\gamma_c = 1\rm ps^{-1} $.
Note that as our results demonstrate, a decrease of the left boundary, $\gamma_m$, even up to $\gamma_m \approx 1 \rm s^{-1}$ practically does not change ET rates. We also introduce the notation: $D=|g_1-g_2|$.

\begin{figure}[tbh]
\begin{center}
\scalebox{0.3}{\includegraphics{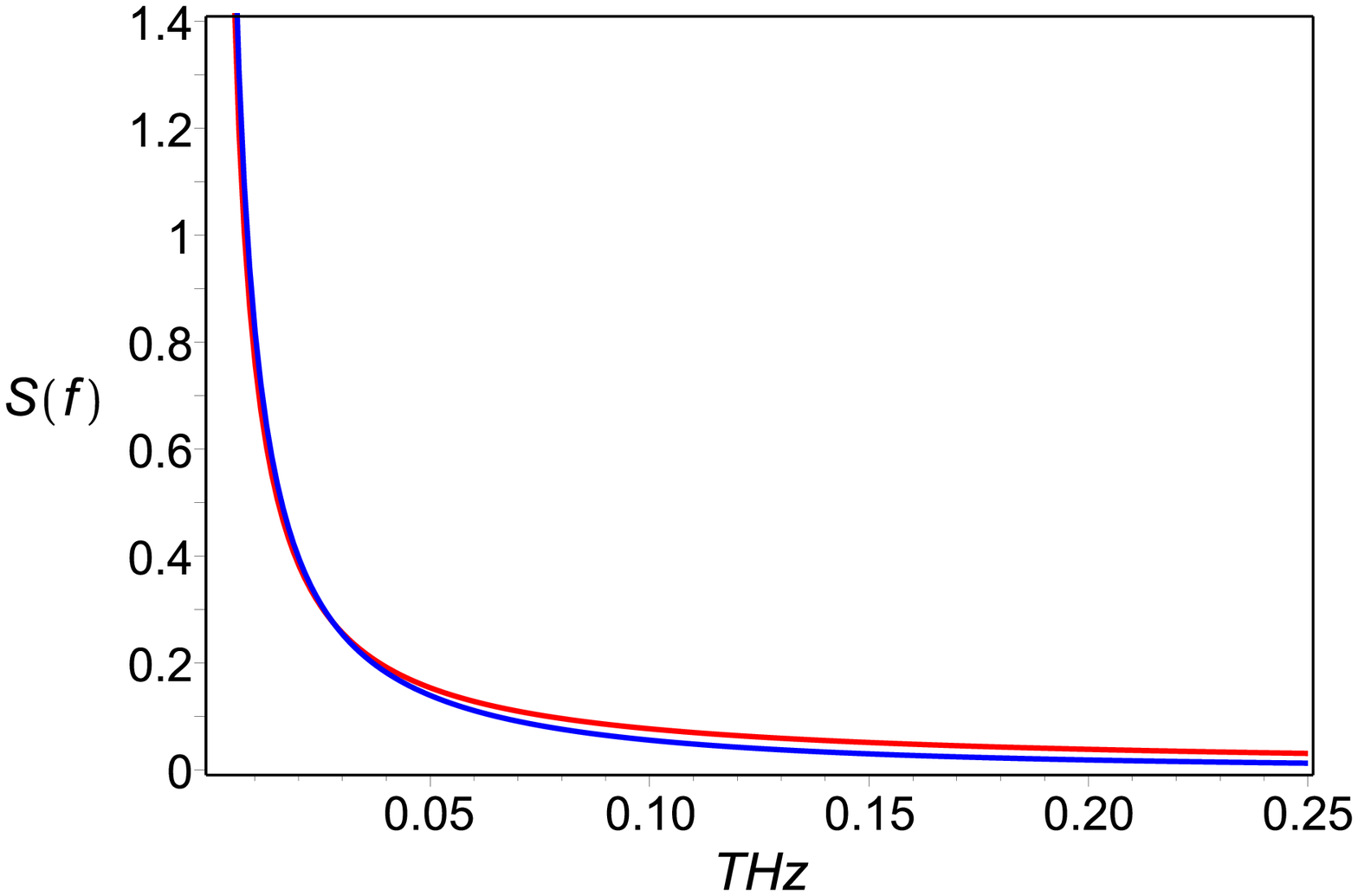}}
\scalebox{0.325}{\includegraphics{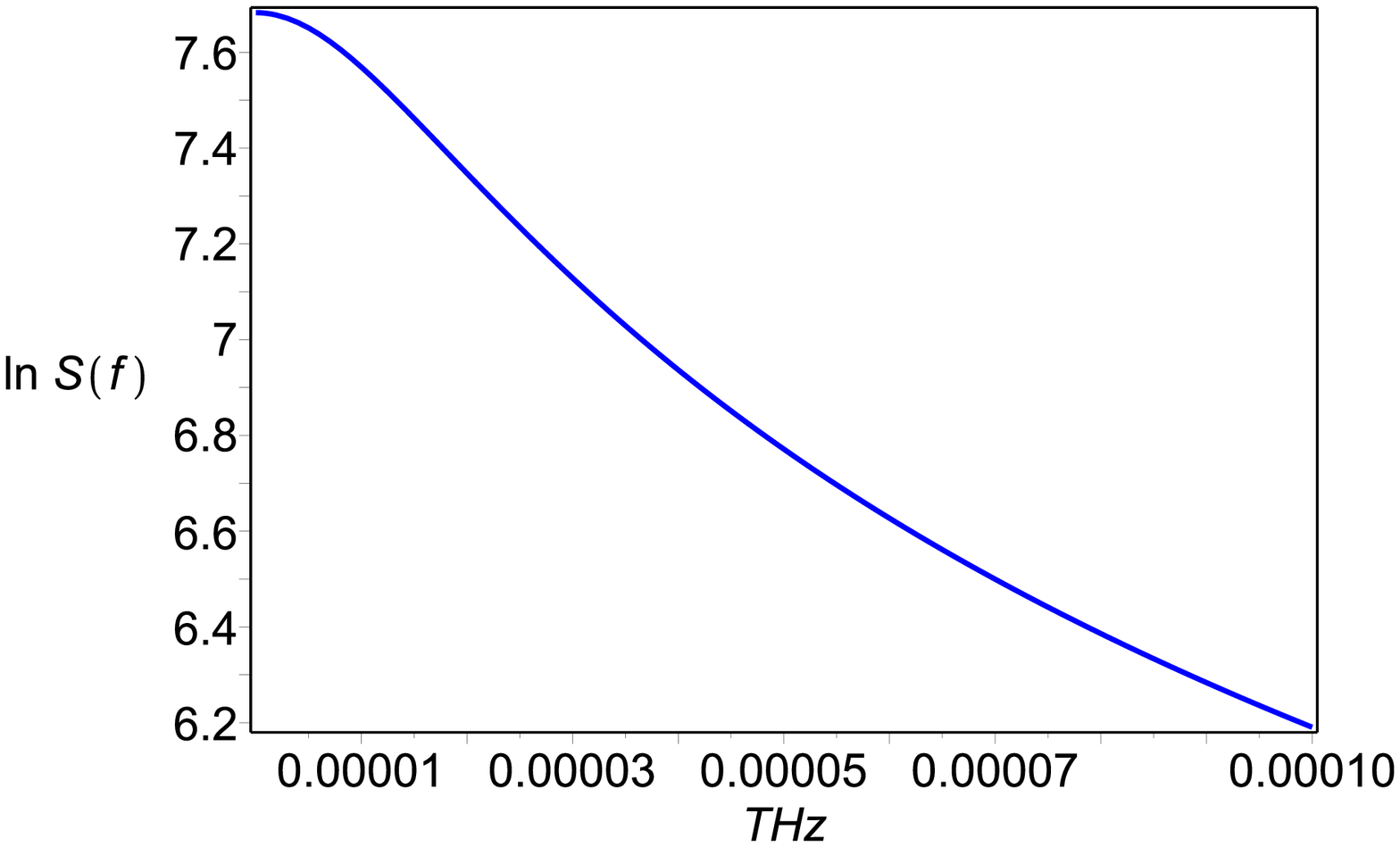}}
\end{center}
\caption{The spectral density of noise $\sigma= 1 $. Left panel: $S(f)$ given by exact formula, Eq.~(\ref{SF1}) (blue line), and the asymptotic formula (\ref{Sf1d}): $S(f) \sim 1/f$ (red line). Right panel: $\ln S(f)$  as a function of frequency.
\label{SP1}}
\end{figure}

The spectral density of noise corresponding to  Eq. (\ref{SF1}) and its asymptotic behavior given by  Eq. (\ref{Sf1d}) is presented in Fig. \ref{SP1}.

\subsection{Influence of noise on the electron transfer rates ($\Gamma = 0$)}

\begin{figure}[tbh]
\begin{center}
\scalebox{0.33}{\includegraphics{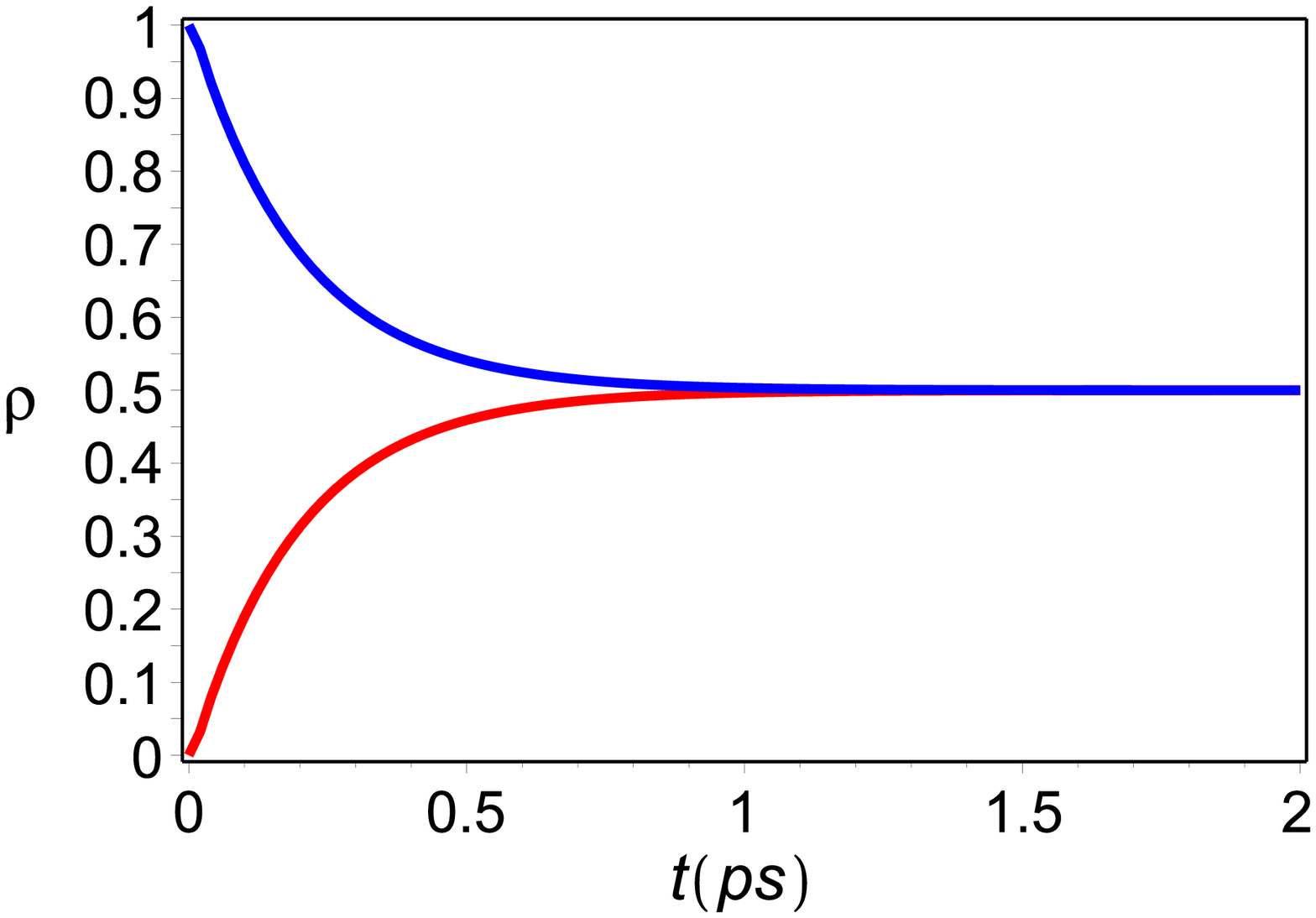}}
\scalebox{0.325}{\includegraphics{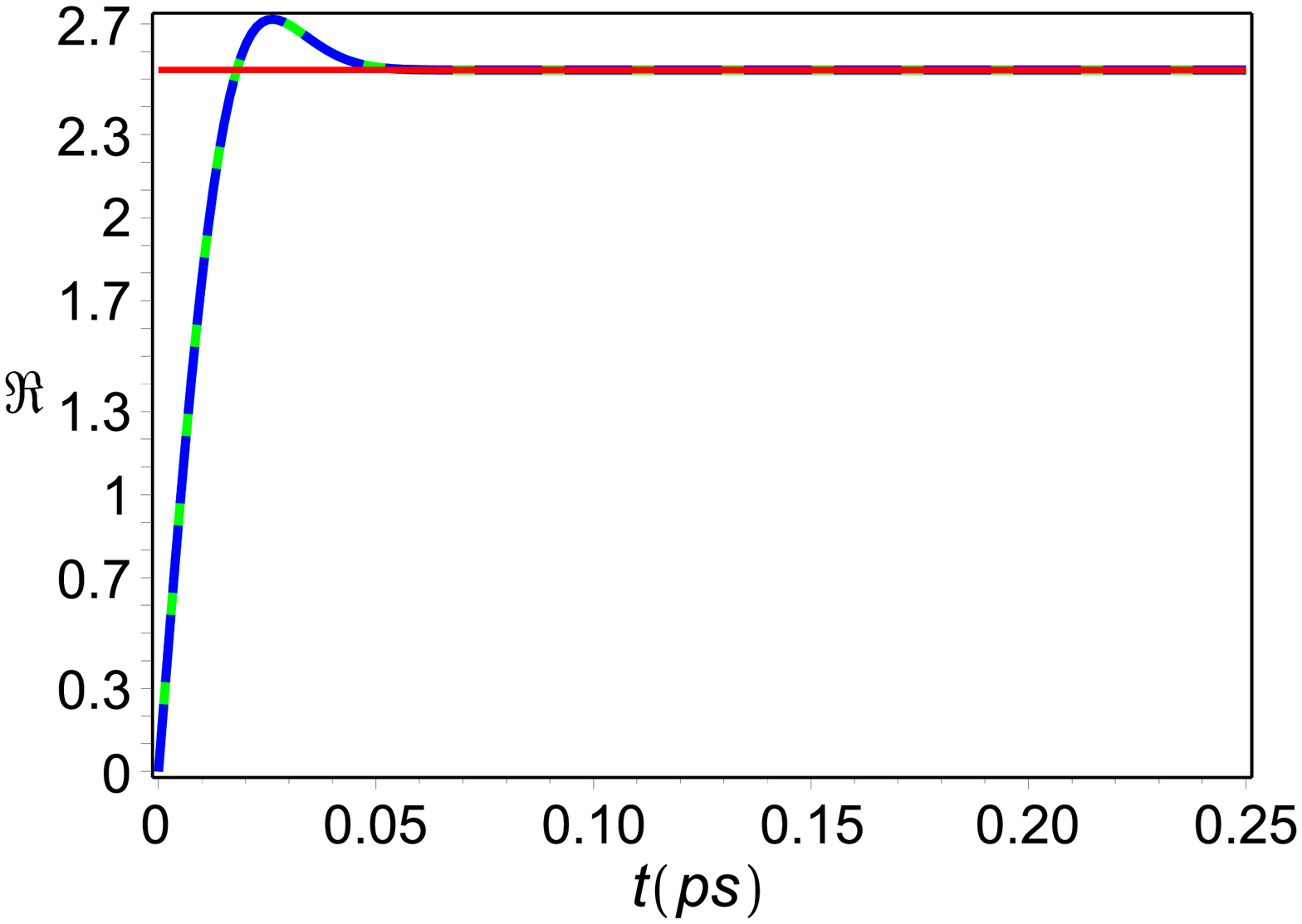}}
\end{center}
\caption{Left panel: Effects of the diagonal noise on the time dependence of the site populations: $\rho_{11}(t)$ (blue line) and $\rho_{22}(t) $ (red line). Right panel: Blue line  describes the time dependence of the ET rate,  $ {\mathfrak R}(t)$, given by Eq.~(\ref{Eq15}), $2\gamma_m = 10^{-4} \rm ps^{-1} $. Green dashed line corresponds to $2\gamma_m = 1 \rm s^{-1}$.
 Red line corresponds to the asymptotic formula (\ref{Eq19f}). The parameters are: $D\sigma=  \varepsilon = 60 \rm ps^{-1} $, $2\gamma_c = 1 \rm ps^{-1} $ and $V = 20 \rm ps^{-1}$.
\label{fast1}}
\end{figure}

For $\Gamma = 0$, we obtain the exact solution of the system (\ref{N4}) and (\ref{N5}),
\begin{eqnarray}\label{Bt}
\big\langle{\rho}_{11}(t)\big\rangle = \frac{1}{2} + \bigg(\rho_{11}(0)-\frac{1}{2}\bigg)\displaystyle e^{-2\int_0^t {\mathfrak R}(t')dt'}, \\
\big\langle{\rho}_{22}(t)\big\rangle = \frac{1}{2} + \bigg(\rho_{22}(0) -\frac{1}{2}\bigg)\displaystyle e^{-2\int_0^t  {\mathfrak R}(t')dt'}. \label{B8}
\end{eqnarray}
It follows that, independent of the initial conditions and the nature of noise (producing decoherence or relaxation), in the limit, $t\rightarrow\infty$, the presence of noise produces equal populations in the two levels. (See Fig. \ref{fast1}, left panel.)

The computation of the ET rate, ${\mathfrak R}(t)$, yields,
\begin{eqnarray}\label{Eq15}
 {\mathfrak R}(t)= \frac{V^2}{4} \int_{-t}^{t}e^{i\varepsilon\tau}\big\langle e^{i\kappa(\tau)} \big\rangle d\tau, \quad
 \kappa(\tau) = -D\int_0^\tau \xi(t')dt'.
\end{eqnarray}
To proceed further, we use the first order cumulant expansion (the Gaussian approximation) to evaluate the generating functional $\big\langle e^{i\kappa(\tau)} \big\rangle$. The computation gives
\begin{eqnarray}
\big\langle e^{i\kappa} \big\rangle =  e^{-\langle\kappa^2\rangle/2} =\exp\bigg(- D^2\int^t_0 dt'\int_{0}^{t'}dt''\chi(t' -t'') \bigg).
\label{Eq15a}
\end{eqnarray}

Let us assume that initially the system occupies only the upper level (donor), $\rho_{11}(0)= 1$. Then, if $g_1 =g_2$ the solution for the diagonal components of the density matrix takes the form,
\begin{eqnarray}\label{Eq18g}
\big\langle{\rho}_{11}(t)\big\rangle = \frac{1}{2} + \frac{1}{2} \exp\Big(-2\frac{V^2  }{\varepsilon^2} \sin^2\frac{\varepsilon t}{2}\Big ), \quad
\big\langle{\rho}_{22}(t)\big\rangle = \frac{1}{2}  -\frac{1}{2}
\exp\Big( -2\frac{V^2  }{\varepsilon^2} \sin^2\frac{\varepsilon t}{2}\Big).
\label{Eq18h}
\end{eqnarray}
One can see that up to the first order in the dimensionless parameter, $V^2/\varepsilon^2$, the approximate solution (\ref{Eq18h}) coincides with the exact solution given by Eq. (\ref{P3d}) (with $\Gamma =0$). In this case, the effect of the collective noise vanishes.

{\em Relation to the Marcus' theory.} The  asymptotic ET rate for $\Gamma =0$ is defined as $ {\mathcal R}= \lim_{t\rightarrow \infty} {\mathfrak R}(t)$. Using Eq.~(\ref{Eq15}), we obtain,
\begin{align} \label{Eq19k}
  {\mathcal R}= \frac{V^2}{4}\int_{-\infty}^{\infty}dt \exp\big(i\varepsilon t   - \Theta(t) \big),
\end{align}
where $\Theta(t) = D^2\int^t_0 dt'\int_{0}^{t'}dt''\chi(t' -t'')$.
To evaluate $\Theta(t)$, we use the approximation, $\chi(t) \approx \chi(0)$. (Note that $\chi(0)= \sigma^2$.) This yields
$\Theta(t) \backsimeq ( \sigma D t)^2/2$. Performing the integration over  $t$ in (\ref{Eq19k}), we obtain
\begin{align}
 {\mathcal R} =\frac{V^2}{4}\sqrt{\frac{2\pi}{D^2 \sigma^2}}\exp\Bigg(-\frac{\varepsilon^2}{2D^2 \sigma^2}\Bigg).
\label{Eq19f}
\end{align}
In Fig. \ref{fast1}, we compare the results of numerical calculations (blue line) of relaxation rate, ${\mathfrak R}(t)$ described by by Eq.~(\ref{Eq15})   with the asymptotic formula (\ref{Eq19f}). One can see a good agreement of the asymptotic rate defined by Eq. (\ref{Eq19f}) with the formula (\ref{Eq15}).

In the case in which the number of thermally excited fluctuators, ${\mathcal N}_T \gg 1 $, the dispersion, $\sigma^2$, is a linear function of the temperature, so that $\sigma^2= P_0kT$ \cite{BGA}. Inserting this expression into Eq.(\ref{Eq19f}), we obtain
\begin{align}
 {\mathcal R}= |V_{12}|^2\sqrt{\frac{\pi}{\lambda  kT}}\exp\Bigg(-\frac{(E_1 -E_2)^2}{4\lambda kT}\Bigg),
\label{Eq19}
\end{align}
where $\lambda = {D^2 P_0}/{2 }$ and $|V_{12}|= V/2$. Comparing our result with the Marcus formula \cite{MRSN}, one can see that the classical noise results in
a large-time asymptotic of the ET rate which can be  expressed in the form of the Marcus-type formula.

\subsection{Noise-assisted electron transfer in the reaction center ($\Gamma \neq 0$)}

\begin{figure}[tbh]
\begin{center}
\scalebox{0.3}{\includegraphics{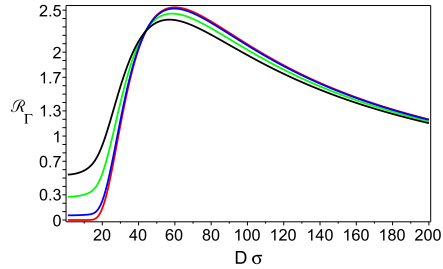}}
\end{center}
\caption{The function ${\mathcal R_\Gamma}$ vs. the amplitude of noise, $D\sigma$ ($\varepsilon =  60 \rm ps^{-1}$, $ V = 20 \rm ps^{-1} $). Black line ($  \Gamma  = 10 \rm ps^{-1} $), green line ($  \Gamma  = 5 \rm ps^{-1} $), blue line ($  \Gamma  = 1 \rm ps^{-1} $), red line ($  \Gamma  = 0 \,\rm ps^{-1} $).
\label{EG}}
\end{figure}

We consider here noise-assisted ET to the sink described by the equations of motion (\ref{N4}) and (\ref{N5}) for the averaged components of the density matrix:
\begin{align} \label{N6a}
\frac{d}{dt}{\langle{\rho}}_{11}(t)\rangle =- {\mathfrak R}(t)\big(\big\langle{\rho}_{11}(t)\big\rangle -\big\langle{\rho}_{22}(t)\big\rangle\big) , \\
\frac{d}{dt}{\langle{\rho}}_{22}(t)\rangle ={\mathfrak R}(t)\big(\big\langle{\rho}_{11}(t)\big\rangle -\big\langle{\rho}_{22}(t)\big\rangle\big) - 2\Gamma \langle {\tilde \rho}_{22}(t)\rangle ,
\label{N7}
\end{align}
where $ {\mathfrak R}(t)= (1/2){V^2} \int_{0}^{t} e^{-\Gamma t}\cos\varepsilon\tau \big\langle e^{i\kappa(\tau)} \big\rangle d\tau$. Further, we use the Gaussian approximation to evaluate the generating functional:
\begin{align}
\big\langle e^{i\kappa(t)} \big\rangle = \exp\bigg(- (g_1 -g_2)^2\int^t_0 dt'\int_{0}^{t'}dt''\chi(t' -t'') \bigg).
\end{align}
Performing the integration over  $t$ in Eq.~(\ref{N6}), we obtain for $\Gamma \ne 0$ a generalization of asymptotic expression for $\mathcal R$ given by Eq.(\ref{Eq19f}) for finite $\Gamma$, ${\mathcal R}_\Gamma= \lim_{t\rightarrow \infty}{\mathfrak R}(t)$, as
\begin{align}
 {\mathcal R}_{\Gamma} =\frac{V^2\sqrt{2\pi}}{8D\sigma} \Bigg( \exp\bigg(\frac{(\Gamma+ i\varepsilon)^2}{2D^2 \sigma^2}\bigg){\rm erfc}\bigg(\frac{\Gamma+ i\varepsilon}{\sqrt{2}D \sigma}\bigg) + \exp\bigg(\frac{(\Gamma - i\varepsilon)^2}{2D^2 \sigma^2}\bigg){\rm erfc}\bigg(\frac{\Gamma - i\varepsilon}{\sqrt{2}D \sigma}\bigg)\Bigg ),
\label{Eq20a}
\end{align}
where ${\rm erf}c(z)$ denotes the complementary error function, ${\rm erfc}(z)= 1 -{\rm erf}(z)$ \cite{abr}. The dependence of ${\mathcal R_\Gamma}$ as function of the amplitude of noise, $D\sigma$, is presented in Fig. \ref{EG}.

Using these results and taking the initial conditions as ${\rho}_{11}(0)=1$, we obtain the approximate solution of Eqs.~(\ref{N6a}) and (\ref{N7})
\begin{align}\label{AS1}
{\langle{\rho}}_{11}(t)\rangle \approx \bigg(\frac{1}{2} - \frac{\Gamma}{2\sqrt{ {\mathcal R}_\Gamma^2 + \Gamma^2}}\bigg)e^{-{\mathcal R}_1t}+  \bigg(\frac{1}{2} + \frac{\Gamma}{2\sqrt{ {\mathcal R}_\Gamma^2 + \Gamma^2}}\bigg)e^{-{\mathcal R}_2t}, \\
{\langle{\rho}}_{22}(t)\rangle \approx  \frac{ {\mathcal R}_\Gamma}{2\sqrt{ {\mathcal R}_\Gamma^2 + \Gamma^2}}\bigg(e^{-{\mathcal R}_2t} -e^{-{\mathcal R}_1t}  \bigg),
\label{AS2}
\end{align}
where ${\mathcal R}_{1,2} =  {\mathcal R}_\Gamma + \Gamma \pm \sqrt{ {\mathcal R}_\Gamma^2 + \Gamma^2} $. Inserting (\ref{AS2}) into Eq. (\ref{Eq16}),  to we obtain for the ET efficiency
\begin{align}\label{Eq21}
\eta(t) = 1- e^{-\frac{({\mathcal R}_1 + {\mathcal R}_2)t}{2}}\bigg( \cosh\frac{({\mathcal R}_1- {\mathcal R}_2)t}{2}  + \frac{{\mathcal R}_1 + {\mathcal R}_2}{{\mathcal R}_1 - {\mathcal R}_2} \sinh\frac{({\mathcal R}_1- {\mathcal R}_2)t}{2} \bigg).
\end{align}
This yields the following asymptotic behavior of the ET efficiency
\begin{align}\label{Eq21a}
\eta(t) \approx 1- \frac{{\mathcal R}_1 }{{\mathcal R}_1 - {\mathcal R}_2} e^{- {\mathcal R}_2 t}.
\end{align}
As can be seen from Eq. (\ref{Eq21}) there are two ET rates, ${\mathcal R}_1$ and ${\mathcal R}_2$. The asymptotic behavior of ET efficiency, $\eta(t)$, is defined by the lowest ET rate, ${\mathcal R}_2$.
\begin{figure}[tbh]
\begin{center}
\scalebox{0.35}{\includegraphics{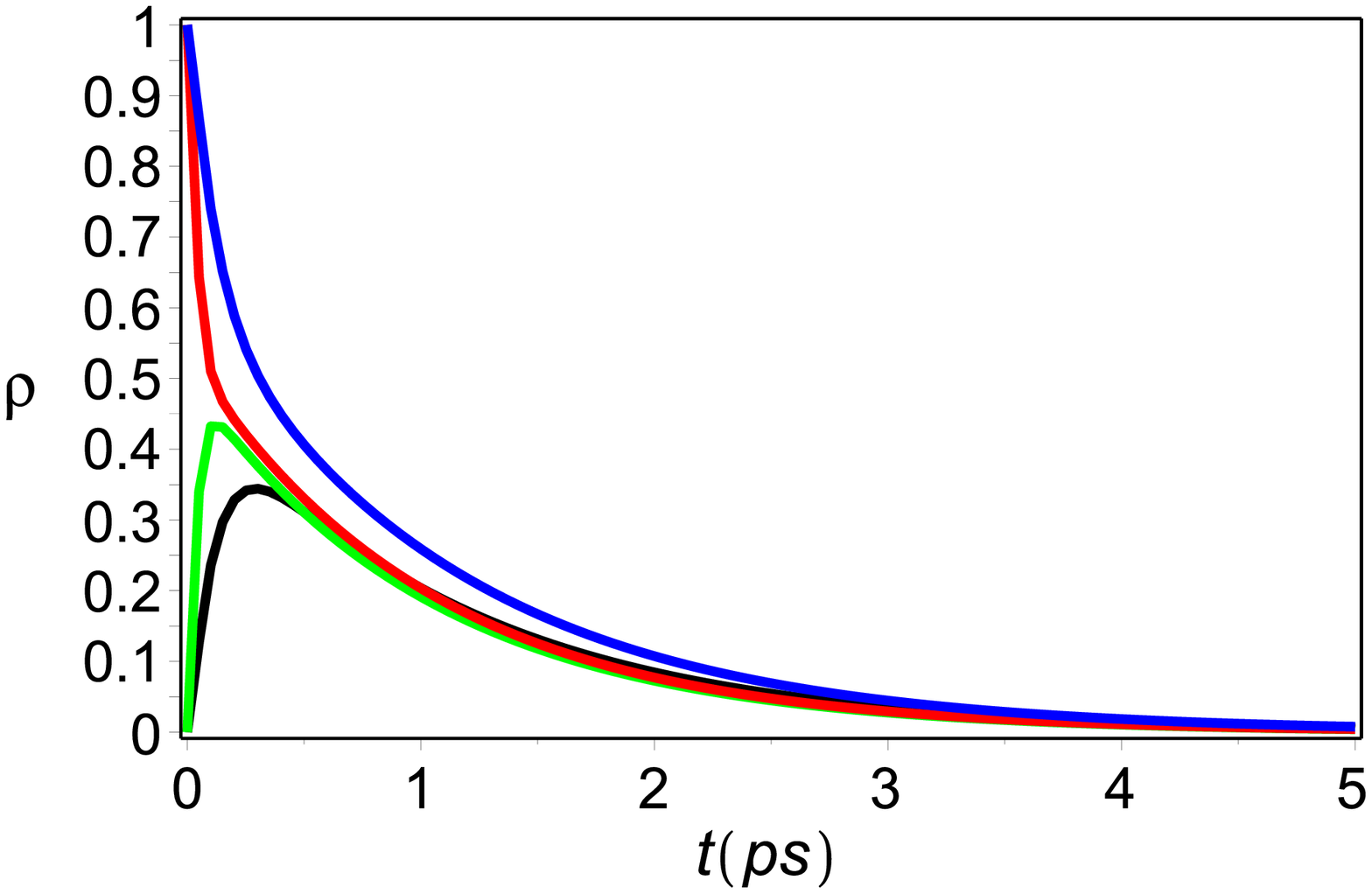}}
\scalebox{0.325}{\includegraphics{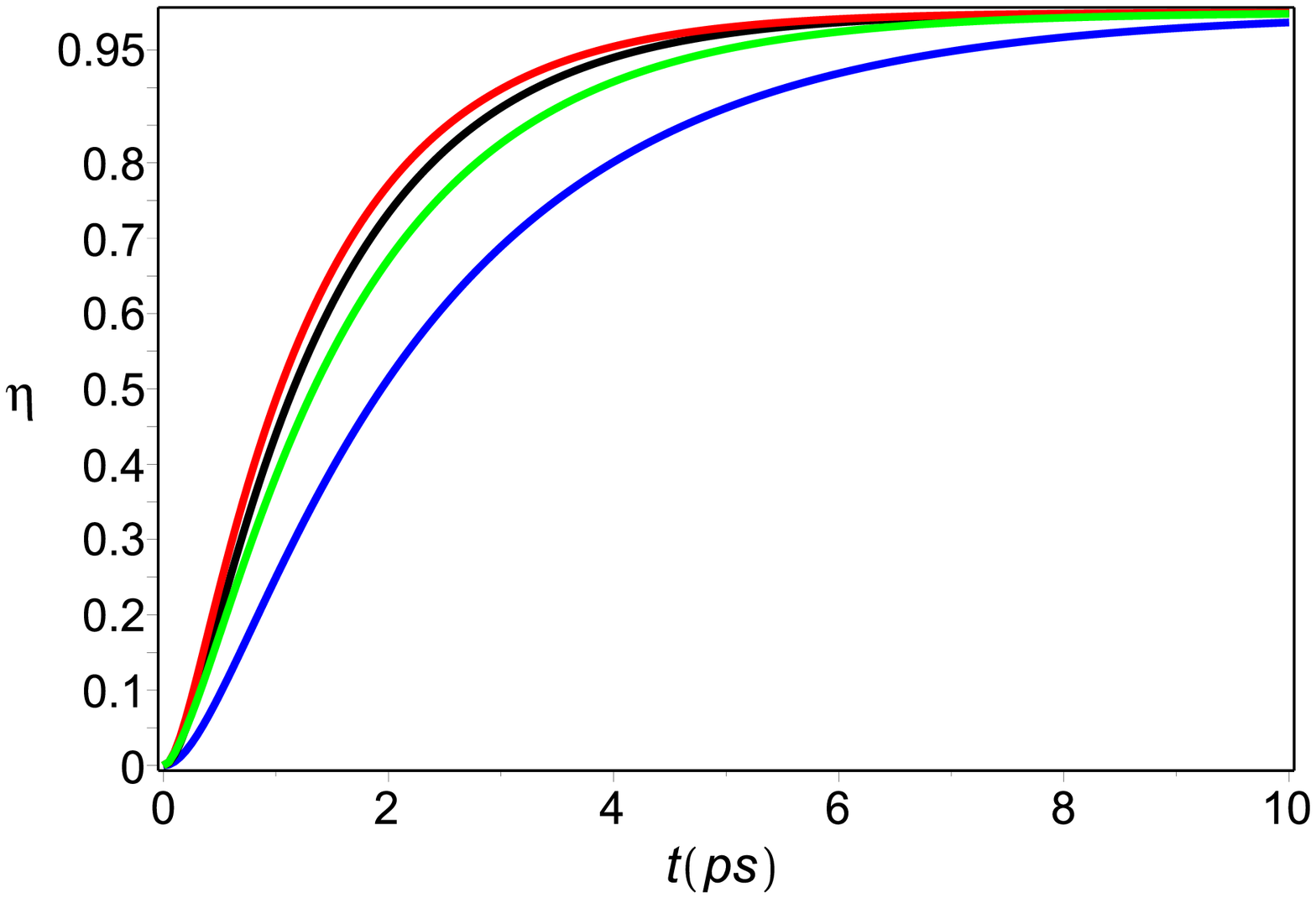}}
\end{center}
\caption{The time dependence of the site population (left panel) and the ET efficiency (right panel) in the presence of noise. Left panel: (i) $V= 20 \rm ps^{-1}$,  blue and black lines correspond to, $\rho_{11}(t) $ and $\rho_{22}(t) $, respectively; (ii) $V= 40 \rm ps^{-1}$, red line  corresponds to $\rho_{11}(t) $, and green line corresponds to $\rho_{22}(t) $. Right panel: The ET efficiency, $\eta(t)$, in the presence of  noise. Green line: ($ V=20 \rm ps^{-1}$, $D\sigma= 30 \rm ps^{-1}$);  red line: ($ V=20 \rm ps^{-1}$, $D\sigma\approx  60 \rm ps^{-1}$); black line: ($ V=20 \rm ps^{-1}$, $D\sigma= 120 \rm ps^{-1}$);  blue line: ($ V=10 \rm ps^{-1}$, $D\sigma = 60 \rm ps^{-1}$). In all cases: $ \varepsilon = 60 \rm ps^{-1}, \Gamma = 1\rm ps^{-1} $, $2\gamma_m = 10^{-4} \rm ps^{-1} $ and $2\gamma_c = 1 \rm ps^{-1} $.
\label{S2}}
\end{figure}

In Fig. \ref{S2}, we present the results of  numerical simulations for $\Gamma =1 \; \rm ps^{-1}$, and for a given sharp redox potential, $\varepsilon=60  \; \rm ps^{-1}$. Red line corresponds to the amplitude of noise which was optimized by using the modified Marcus-type formula given by Eq. (\ref{Eq20a}). (See Fig. \ref{EG}.)  As one can see from a comparison of the results presented in Fig.~\ref{R1} (for a sharp redox potential) and Fig.~\ref{S2}, influence of noise with amplitudes near to optimal value significantly accelerates the ET to the sink.

\section{Conclusions}

In this paper, we model quantum electron transfer dynamics in a photosynthetic reaction center consisting of  three elementary pigment units. Two of them, a donor and an acceptor, are represented by localized sites of protein pigments  with discrete energy levels. The donor interacts with the acceptor  through the corresponding matrix element. The third protein pigment (sink) has a continuous energy spectrum, and is described by two parameters: its density of states and its strength of interaction with the acceptor. The sink is described self-consistently, by using the Feshbach projection method on the ``donor-acceptor" intrinsic states, within a non-Hermitian Hamiltonian approach. We apply our results to the quantum dynamics of the electron transfer in the active branch of the  quinone-type PSII reaction center.

The collective external noise produced by the environment of the proteins acts on the ``donor-acceptor" sub-system. Usually, the presence of noise acts as an incoherent pump in the system under consideration. But, as our results demonstrate, the simultaneous influence of both noise and the sink, significantly assist, under appropriate conditions, the quantum efficiency of the electron transfer. We derived the expression for electron transfer rate which describes the tunneling to the sink, in the presence of noise. We calculate explicitly the corresponding region of parameters of the noise assisted quantum electron transfer for sharp and flat redox potentials, and for noise described by an ensemble of two-level fluctuators.

Our results show that even in this simplified model, the quantum dynamics of the electron transfer to the sink can be rather  complicated, and depends on many parameters. Further analytical research and numerical simulations are required to extend our approach for (i) complicated dependencies of the density of states on energy in the sinks for  flat and sharp redox potentials,  in the presence of noise and thermal environments, and (ii) more complicated LHCs-RCs complexes. The problem of the electron transfer optimization also requires further analysis.

\section*{Acknowledgements}

We are thankful to B.H. McMahon for useful discussions. This work was carried out under the auspices of the National Nuclear Security Administration of the U.S. Department of Energy at Los Alamos National Laboratory under Contract No.
DE-AC52-06NA25396. A.I. Nesterov acknowledges the support from the CONACyT, Grant No. 118930.

\appendix

\section{Solution of the Liouville equation}

The solution of the the Liouville equation (\ref{DM1}),
$i\dot{ \rho} = [\mathcal H,\rho] - i \{\mathcal W,\rho\}$,
is given by\cite{NAI,NBB},
\begin{align}\label{P3a}
\rho_{11}(t)  = {e^{-\Gamma t}} \bigg |\Big(\cos\frac{\Omega t}{2} - i\cos\theta\sin\frac{\Omega t}{2}\Big) C_1-
i \sin\theta\sin\frac{\Omega t}{2} C_2\bigg |^2,\\
\rho_{22}(t)  = {e^{-\Gamma t}} \bigg |\Big(\cos\frac{\Omega t}{2} +i\cos\theta\sin\frac{\Omega t}{2}\Big) C_2-
i \sin\theta\sin\frac{\Omega t}{2} C_1\bigg |^2, \label{P3b}
\end{align}
\begin{align}
\rho_{12}(t)  = {e^{-\Gamma t}} \bigg (\Big(\cos\frac{\Omega t}{2} - i\cos\theta\sin\frac{\Omega t}{2}\Big) C_1-
i \sin\theta\sin\frac{\Omega t}{2} C_2\bigg )^\ast\cdot \nonumber \\
\bigg (\Big(\cos\frac{\Omega t}{2} + i\cos\theta\sin\frac{\Omega t}{2}\Big) C_2-
i \sin\theta\sin\frac{\Omega t}{2} C_1\bigg ) , \\
\rho_{21}(t)  = {e^{-\Gamma t}} \bigg (\Big(\cos\frac{\Omega t}{2} - i\cos\theta\sin\frac{\Omega t}{2}\Big) C_1-
i \sin\theta\sin\frac{\Omega t}{2} C_2\bigg )\cdot \nonumber \\
\bigg (\Big(\cos\frac{\Omega t}{2} + i\cos\theta\sin\frac{\Omega t}{2}\Big) C_2- i \sin\theta\sin\frac{\Omega t}{2} C_1\bigg )^\ast,
\end{align}
where $\cos\theta = (\varepsilon + i\Gamma )/\Omega$, $\sin\theta = V/\Omega$, and  $\Omega= \sqrt{V^2 +(\varepsilon + i\Gamma )^2}$ is the complex Rabi frequency. The constants, $C_1$ and $C_2$, are defined from the initial conditions as follows:  $\rho_{11}(0)=|C_1|^2$, $\rho_{22}(0)=|C_2|^2$,  $\rho_{12}(0)=C^\ast_1 C_2$ and  $\rho_{21}(0)=C^\ast_2 C_1$.

Let us assume that initially only the acceptor site is occupied, so that $\rho_{11}(0)=0$ and $\rho_{22}(0)= 1$. This yields,
\begin{align}
\rho_{11}(t)  = {e^{-\Gamma t}} \bigg |\sin\theta\sin\frac{\Omega t}{2}\bigg |^2, \quad
\rho_{22}(t)  = {e^{-\Gamma t}} \bigg |\Big(\cos\frac{\Omega t}{2} + i\cos\theta\sin\frac{\Omega t}{2}\Big)\bigg|^2, \\
\rho_{12}(t)  = i{e^{-\Gamma t}} \bigg (\sin\theta\sin\frac{\Omega t}{2} \bigg )^\ast\cdot \bigg (\cos\frac{\Omega t}{2} + i\cos\theta\sin\frac{\Omega t}{2} \bigg ) , \\
\rho_{21}(t)  = -i{e^{-\Gamma t}} \bigg (\sin\theta\sin\frac{\Omega t}{2} \bigg )\cdot \bigg (\cos\frac{\Omega t}{2} + i\cos\theta\sin\frac{\Omega t}{2} \bigg )^\ast .
\end{align}
Taking $V=0$, we find $\rho_{11}(t)=\rho_{12}(t)= \rho_{21}(t)=0$, and for $\rho_{22}(t) $ we obtain the Weisskopf-Wigner formula for an irreversible decay, $\rho_{22}(t)  = {e^{-2\Gamma t}} $ \cite{WW,SM}.

Presenting $\Omega= \Omega_1 +i \Omega_2 = \sqrt{p +i q}$, where
$p=  V^2+ \varepsilon^2- \Gamma^2$ and $q= 2\varepsilon\Gamma$, we obtain
$\Omega_1^2 - \Omega_2^2 = p$ and $\Omega_1\Omega_2 = q$.
This yields, $\Omega_1= \pm\frac{1}{\sqrt{2}} \sqrt{p + \sqrt{p^2 + q^2}}, \quad
\Omega_2= \pm\frac{1}{\sqrt{2}} \sqrt{-p + \sqrt{p^2 + q^2}}$,
where the upper sign corresponds to $p>0$, and the lower sign corresponds to $p<0$. Using these results, we obtain for $\rho_{22}(t) $ the simple analytical expression,
\begin{align}
\rho_{22}(t)  = \frac{  V^2 e^{-\Gamma t}}{2(\Omega^2_1 +\Omega^2_2)} \big (\cosh{\Omega_2 t} - \cos{\Omega_1 t} \big ).
\label{P3e}
\end{align}

\section{Equation of motion for the average density matrix}

In the interaction representation, considering the off-diagonal elements as perturbations, so that $\tilde{\mathcal H}={\mathcal H}_0 + V(t)- i\mathcal W$, where
\begin{align}
 {\mathcal H}_0= \sum_{n} \varepsilon_n |n\rangle\langle  n | +  \sum_{n} \lambda_{nn} (t) |n\rangle\langle  n | , \\
V(t)= \sum_{m \neq n} ( V_{mn} +\lambda_{mn}(t))|m\rangle\langle  n |, \quad \mathcal W = \Gamma |2\rangle \langle 2|,
\end{align}
we obtain the following equations of motion,
\begin{align}  \label{A1a}
{\dot {\tilde \rho}}_{11} = i({\tilde \rho}_{12}{\tilde V}_{21}- {\tilde V}_{12} {\tilde \rho}_{21}), \quad
{\dot {\tilde \rho}}_{22} = i({\tilde \rho}_{21}{\tilde V}_{12}- {\tilde V}_{21} {\tilde \rho}_{12})-2\Gamma  {\tilde \rho}_{22},  \\
{\dot {\tilde \rho}}_{12} = i{\tilde V}_{12}({\tilde \rho}_{11}-  {\tilde \rho}_{22})- \Gamma  {\tilde \rho}_{12}, \quad
{\dot {\tilde \rho}}_{21} = i{\tilde V}_{21}({\tilde \rho}_{11}-  {\tilde \rho}_{22}) -  \Gamma  {\tilde \rho}_{21},
\label{A1b}
\end{align}
where $ \tilde \rho= T(e^{i\int_0^t H_0(\tau) d \tau})\rho T(e^{-i\int_0^t H_0(\tau) d\tau })$ and $ \tilde V=  T(e^{i\int_0^t H_0(\tau) d \tau})V T(e^{-i\int_0^t H_0(\tau) d\tau })$.

Using Eqs. (\ref{A1a})- (\ref{A1b}), we obtain
\begin{align} \label{Eq2}
 {\tilde \rho}_{11}(t) = { {\tilde \rho}}_{11}(0) + i\int_0^t({\tilde \rho}_{12}(t'){\tilde V}_{21}(t')- {\tilde V}_{12}(t') {\tilde \rho}_{21}(t'))dt', \\
 {\tilde \rho}_{22}(t) =  {\tilde \rho}_{22}(0)+ i\int_0^t e^{-2\Gamma (t- t')}({\tilde \rho}_{21}(t'){\tilde V}_{12}(t')- {\tilde V}_{21}(t') {\tilde \rho}_{12}(t')), \\
 {\tilde \rho}_{12}(t) = {\tilde \rho}_{12}(0) + i \int_0^t e^{-\Gamma (t- t')}{\tilde V}_{12}(t') ({\tilde \rho}_{11}(t')-  {\tilde \rho}_{22}(t'))dt', \\
{\tilde \rho}_{21}(t) = {\tilde \rho}_{21}(0) + i  \int_0^t e^{-\Gamma (t- t')}{\tilde V}_{21}(t')({\tilde \rho}_{11}(t')-  {\tilde \rho}_{22}(t'))dt'.
\label{Eq3a}
\end{align}
We assume that initially ${\tilde \rho}_{12}(0)={\tilde \rho}_{21}(0)=0$. Now inserting  (\ref{Eq2}) - (\ref{Eq3a}) into Eqs. (\ref{A1a}) - (\ref{A1b}), and taking into account that ${\tilde \rho}_{11} =  \rho_{11}$ and ${\tilde \rho}_{22} =  \rho_{22}$,  we obtain the following system of integro-differential equations,
\begin{align}  \label{B3}
&{\dot {\rho}}_{11} = - \int_0^t e^{-\Gamma (t- t')}\Big({\tilde V}_{21}(t){\tilde V}_{12}(t')+ {\tilde V}_{21}(t'){\tilde V}_{12}(t)\Big)\Big({ \rho}_{11}(t') -{ \rho}_{22}(t')\Big) dt', \\
&{\dot {\rho}}_{22} =  \int_0^te^{-\Gamma (t- t')}\Big({\tilde V}_{21}(t){\tilde V}_{12}(t')+ {\tilde V}_{21}(t'){\tilde V}_{12}(t)\Big)\Big({ \rho}_{11}(t') -{\rho}_{22}(t')\Big) dt' - 2\Gamma {\rho}_{22}(t) , \\
&\dot{\tilde \rho}_{12}(t) =  - \int_0^t\Big(1+ e^{-2\Gamma (t- t')}\Big) \Big({\tilde V}_{21}(t'){\tilde \rho}_{12}(t')-  {\tilde V}_{12}(t'){\tilde \rho}_{21}(t')\Big ){\tilde V}_{12}(t)dt'- \Gamma  {\rho}_{12}(t) \nonumber \\
&+ i{\tilde V}_{12}(t)({\tilde \rho}_{11}(0)-  {\tilde \rho}_{22}(0))  , \\
&\dot{\tilde \rho}_{21}(t) = - \int_0^t\Big(1+ e^{-2\Gamma (t- t')}\Big) \Big({\tilde V}_{21}(t'){\tilde \rho}_{12}(t')-  {\tilde V}_{12}(t'){\tilde \rho}_{21}(t')\Big ){\tilde V}_{21}(t)dt'- \Gamma  {\rho}_{21}(t) \nonumber \\
&+ i{\tilde V}_{21}(t)({\tilde \rho}_{11}(0)-  {\tilde \rho}_{22}(0)) .
 \label{B4}
\end{align}
For the average components of the density matrix this yields,
\begin{align}  \label{Eq3}
&\frac{d}{dt}{\langle{\rho}}_{11}(t)\rangle  = - \int_0^te^{-\Gamma (t- t')}\Big\langle\Big({\tilde V}_{21}(t){\tilde V}_{12}(t')+ {\tilde V}_{21}(t'){\tilde V}_{12}(t)\Big)\Big({ \rho}_{11}(t') -{ \rho}_{22}(t')\Big) \Big\rangle dt', \\
&\frac{d}{dt}{\langle{\rho}}_{22}(t)\rangle  =  \int_0^te^{-\Gamma (t- t')}\Big\langle\Big({\tilde V}_{21}(t){\tilde V}_{12}(t')+ {\tilde V}_{21}(t'){\tilde V}_{12}(t)\Big)\Big({ \rho}_{11}(t') -{\rho}_{22}(t')\Big)\Big\rangle dt' \nonumber \\
&- 2\Gamma \langle{\rho}_{22}(t)\rangle , \\
&\frac{d}{dt}{\langle{\rho}}_{12}(t)\rangle  =  - \int_0^t\Big(1+ e^{-2\Gamma (t- t')}\Big)\Big\langle \Big({\tilde V}_{21}(t'){\tilde \rho}_{12}(t')-  {\tilde V}_{12}(t'){\tilde \rho}_{21}(t')\Big ){\tilde V}_{12}(t)\Big\rangle dt'- \Gamma  \langle{\rho}_{12}(t)\rangle \nonumber \\
&+ i\langle{\tilde V}_{12}(t)\rangle({ \rho}_{11}(0)-  { \rho}_{22}(0))  , \\
&\frac{d}{dt}{\langle{\rho}}_{21}(t)\rangle  = - \int_0^t\Big(1+ e^{-2\Gamma (t- t')}\Big)\Big\langle \Big({\tilde V}_{21}(t'){\tilde \rho}_{12}(t')-  {\tilde V}_{12}(t'){\tilde \rho}_{21}(t')\Big ){\tilde V}_{21}(t)\Big\rangle dt'- \Gamma \langle {\rho}_{21}(t)\rangle \nonumber \\
&+ i\langle{\tilde V}_{21}(t)\rangle({\rho}_{11}(0)-  {\rho}_{22}(0)) ,
 \label{Eq4}
\end{align}
where the average $\langle\; \rangle$ is taken over the random process describing noise. Generalization of the obtained results for the case ${\tilde \rho}_{12}(0) \neq 0$ and ${\tilde \rho}_{21}(0)\neq 0$ is straightforward.

In the spin-fluctuator model of noise with the number of fluctuators ${\mathcal N} \gg 1$ one has the following relations for the splitting of correlations \cite{NB1b},
\begin{align}
\big\langle\big({\tilde V}_{21}(t){\tilde V}_{12}(t')+ {\tilde V}_{21}(t'){\tilde V}_{12}(t)\big)\big({\tilde\rho}_{11}(t') -{\tilde\rho}_{22}(t')\big)\big \rangle = \nonumber \\
\big(\big\langle{\tilde V}_{21}(t){\tilde V}_{12}(t') \big\rangle + \big\langle{\tilde V}_{21}(t'){\tilde V}_{12}(t)\big\rangle\big)\big(\big\langle{\tilde\rho}_{11}(t')\big\rangle -\big\langle{\tilde\rho}_{22}(t')\big\rangle\big),
\end{align}
and so on. Next, using the second-order cumulant expansion, we obtain the following system of differential equations for the average components of the density matrix,
\begin{align} \label{Eq8a}
&\frac{d}{dt}{\langle{\rho}}_{11}(t)\rangle =-\int_0^t e^{-\Gamma (t- t')}\big(\big\langle{\tilde V}_{21}(t){\tilde V}_{12}(t') \big\rangle + \big\langle{\tilde V}_{21}(t'){\tilde V}_{12}(t)\big\rangle\big)dt'\big(\big\langle{\rho}_{11}(t)\big\rangle -\big\langle{\rho}_{22}(t)\big\rangle\big) \nonumber \\
& + {\mathcal O}(V^4), \\
&\frac{d}{dt}{\langle{\rho}}_{22}(t)\rangle =\int_0^t e^{-\Gamma (t- t')}\big(\big\langle{\tilde V}_{21}(t){\tilde V}_{12}(t') \big\rangle + \big\langle{\tilde V}_{21}(t'){\tilde V}_{12}(t)\big\rangle\big)dt'\big(\big\langle{\rho}_{11}(t)\big\rangle -\big\langle{\rho}_{22}(t)\big\rangle\big) \nonumber \\
&- 2\Gamma \langle {\tilde \rho}_{22}(t)\rangle + {\mathcal O}(V^4),
\end{align}
\begin{align}
&\frac{d}{dt}\langle {\tilde \rho}_{12}(t)\rangle =  i\langle{\tilde V}_{12}(t)\rangle({ \rho}_{11}(0)-  { \rho}_{22}(0)) - \int_0^t \Big(1+ e^{-2\Gamma (t- t')}\Big) \langle{\tilde V}_{12}(t) {\tilde V}_{21}(t')\rangle dt'\langle{\tilde \rho}_{12}(t)\rangle \nonumber \\
&+  \int_0^t \Big(1+ e^{-2\Gamma (t- t')}\Big) \langle{\tilde V}_{12}(t) {\tilde V}_{12}(t')\rangle dt'\langle{\tilde \rho}_{21}(t)\rangle   - \Gamma  \langle{\rho}_{12}(t)\rangle + {\mathcal O}(V^4),\\
&\frac{d}{dt}\langle {\tilde \rho}_{21}(t)\rangle =  i\langle{\tilde V}_{21}(t)\rangle({ \rho}_{11}(0)-  { \rho}_{22}(0))
- \int_0^t \Big(1+ e^{-2\Gamma (t- t')}\Big) \langle{\tilde V}_{21}(t) {\tilde V}_{21}(t')\rangle dt'\langle{\tilde \rho}_{12}(t)\rangle  \nonumber \\
& +  \int_0^t \Big(1+ e^{-2\Gamma (t- t')}\Big)\langle{\tilde V}_{21}(t) {\tilde V}_{12}(t')\rangle dt'\langle{\tilde \rho}_{21}(t)\rangle  - \Gamma  \langle{\rho}_{21}(t)\rangle + {\mathcal O}(V^4).
\end{align}
We rewrite the equation of motion for the diagonal components of the density matrix as,
\begin{align} \label{B4}
\frac{d}{dt}{\langle{\rho}}_{11}(t)\rangle =- {\mathfrak R}(t)\big(\big\langle{\rho}_{11}(t)\big\rangle -\big\langle{\rho}_{22}(t)\big\rangle\big), \\
\frac{d}{dt}{\langle{\rho}}_{22}(t)\rangle ={\mathfrak R}(t)\big(\big\langle{\rho}_{11}(t)\big\rangle -\big\langle{\rho}_{22}(t)\big\rangle\big) - 2\Gamma \langle {\tilde \rho}_{22}(t)\rangle,
\label{B5}
\end{align}
where ${\mathfrak R}(t) =\int_0^t e^{-\Gamma (t- t')}\big(\big\langle{\tilde V}_{21}(t){\tilde V}_{12}(t') \big\rangle + \big\langle{\tilde V}_{21}(t'){\tilde V}_{12}(t)\big\rangle\big)dt'$.

\providecommand{\WileyBibTextsc}{}
\let\textsc\WileyBibTextsc
\providecommand{\othercit}{}
\providecommand{\jr}[1]{#1}
\providecommand{\etal}{~et~al.}


\begin{thebibliography}{[10]}

\othercit
\bibitem{BER1}
 \textsc{R.~Blankenship},
{Molecular} {Mechansms} of {Photosynthesus} (World Scientific, London, 2002).


\bibitem{GSR}
 \textsc{T.~Gustafson} and  \textsc{R.~Sayre},
 \jr{PNAS USA} \textbf{99}, 4091 (2002).


\bibitem{XSG}
 \textsc{L.~Xiong},  \textsc{M.~Seibert},  \textsc{A.~Gusev},
  \textsc{M.~Wasielewski},  \textsc{C.~Hemann},  \textsc{C.~Hille},  and
  \textsc{R.~Sayre},
 \jr{J. Phys. Chem. B} \textbf{108}, 16904 (2004).


\bibitem{Psr}
 \textsc{Z.~Perrine} and  \textsc{R.~Sayre},
 \jr{Biochemistry} \textbf{50}, 1454 (2011).


\bibitem{KHM}
 \textsc{J.~Koepke},  \textsc{X.~Hu},  \textsc{C.~Münke},
  \textsc{K.~Schulten},  and  \textsc{H.~Michel},
 \jr{Structure} \textbf{4}, 581 (1996).


\bibitem{HDR}
 \textsc{X.~Hu},  \textsc{A.~Damjanovic},  \textsc{T.~Ritz},  and
  \textsc{K.~Schulten},
 \jr{Proc. Natl. Acad. Sci. USA} \textbf{95}, 5935 (1998).


\bibitem{CWW}
 \textsc{E.~Collini},  \textsc{C.~Wong},  \textsc{K.~Wilk},  \textsc{P.~Curmi},
   \textsc{P.~Brumer},  and  \textsc{G.~Scholes},
 \jr{Nature Letters} \textbf{463}, 644 (2010).


\bibitem{MRSN}
 \textsc{R.~Marcus} and  \textsc{N.~Sutin},
 \jr{Biochimica et Biophysica Acta} \textbf{811}, 265 (1985).


\bibitem{XSK}
 \textsc{D.~Xu} and  \textsc{K.~Schulten},
 \jr{Chemical Physics} \textbf{182}, 91 (1994).


\bibitem{SWB}
 \textsc{S.~Skourtis},  \textsc{D.~Waldeck},  and  \textsc{D.~Beratan},
 \jr{Annu. Rev. Phys. Chem.} \textbf{64}, 461 (2010).


\othercit
\bibitem{MFL}
 \textsc{B.~McMahon},  \textsc{P.~Fenimore},  and  \textsc{M.~LaBute},
{Protein noises},
 in: {Fluctuations and Noise in Biological, Biophysical, and Biomedical
  Systems}, edited by S.\,M. Bezrukov, H.~Frauenfelder,  and F.~Moss,
  Proceedings of SPIE Vol.\,5110 (2003),  pp.\,10 -- 21.


\bibitem{BPM}
 \textsc{A.~Bar-Even},  \textsc{J.~Paulsson},  \textsc{N.~Maheshri},
  \textsc{M.~Carmi},  \textsc{E.~O'Shea},  \textsc{Y.~Pilpel},  and
  \textsc{N.~Barkai},
 \jr{Nature Genetics} \textbf{38}, 636 (2006).


\bibitem{DBJ}
 \textsc{T.~Dewey} and  \textsc{J.~Bann},
 \jr{Biophys. J.} \textbf{63}, 594 (1992).


\bibitem{NB1b}
 \textsc{A.\,I. Nesterov} and  \textsc{G.\,P. Berman},
 \jr{arXiv:1201.2316 [quant-ph]} (2011).


\bibitem{ECR}
 \textsc{G.~Engel},  \textsc{T.~Calhoun},  \textsc{E.~Read},  \textsc{T.~Ahn},
  \textsc{T.Mancal},  \textsc{Y.~Cheng},  \textsc{R.~Blankenship},  and
  \textsc{G.~Fleming},
 \jr{Nature Letters} \textbf{446}, 782 (2007).


\bibitem{CWWC}
 \textsc{E.~Collini},  \textsc{C.~Wong},  \textsc{K.~Wilk},  \textsc{P.~Curmi},
   \textsc{P.~Brumer},  and  \textsc{G.~Scholes},
 \jr{Nature Letters} \textbf{463}, 644 (2010).


\bibitem{PHFC}
 \textsc{G.~Panitchayangkoon},  \textsc{D.~Hayes},  \textsc{K.~Fransted},
  \textsc{J.~Caram},  \textsc{E.~Harel},  \textsc{J.~Wenb},
  \textsc{R.~Blankenship},  and  \textsc{G.~Engel},
 \jr{PNAS USA} \textbf{107}, 12766 (2010).


\bibitem{IFG}
 \textsc{A.~Ishizaki} and  \textsc{G.~Fleming},
 \jr{PNAS USA} \textbf{106}, 17255 (2009).


\bibitem{RMKL}
 \textsc{P.~Rebentrost},  \textsc{M.~Mohseni},  \textsc{I.~Kassal},
  \textsc{S.~Lloyd},  and  \textsc{A.~Aspuru-Guzik},
 \jr{New Journal of Physics} \textbf{11}(3), 033003 (2009).


\bibitem{CFMB}
 \textsc{G.~Celardo},  \textsc{F.~Borgonovi},  \textsc{M.~Merkli},
  \textsc{V.~Tsifrinovich},  and  \textsc{G.~Berman},
 \jr{arXiv:1111.5443v1. [cond-mat.]} (2011).


\bibitem{PPM}
 \textsc{R.~Pin\ifmmode\,\check{c}\else \v{c}\fi{}\'ak} and
  \textsc{M.~Pudlak},
 \jr{Phys. Rev. E} \textbf{64}, 031906 (2001).


\bibitem{AMS}
 \textsc{D.~Abramavicius} and  \textsc{S.~Mukamel},
 \jr{J. Phys. Chem.} \textbf{133}, 184501 (2010).


\bibitem{YHNN}
 \textsc{F.~Yoshihara},  \textsc{K.~Harrabi},  \textsc{A.\,O. Niskanen},
  \textsc{Y.~Nakamura},  and  \textsc{J.\,S. Tsai},
 \jr{Phys. Rev. Lett.} \textbf{97}(16), 167001 (2006).


\bibitem{THA}
 \textsc{A.~Thilagam},
 \jr{J. Chem. Phys.} \textbf{136}, 065104 (2012).


\bibitem{RI}
 \textsc{I.~Rotter},
 \jr{Phys. Rev. E} \textbf{64}, 036213 (2001).


\bibitem{RI2}
 \textsc{I.~Rotter},
 \jr{Reports on Progress in Physics} \textbf{54}, 635 (1991).


\bibitem{RI3}
 \textsc{I.~Rotter},
 \jr{J. Phys. A} \textbf{42}, 153001 (2009).


\bibitem{VZ}
 \textsc{A.~Volya} and  \textsc{V.~Zelevinsky},
 \jr{Phys. Rev. Lett.} \textbf{94}, 052501 (2005).


\othercit
\bibitem{SM}
 \textsc{S.~Mukamel},
{Principles} of {Nonlinear} {Optical} {Spectroscopy} (Oxford University Press,
  New York, 1995).


\bibitem{LVD}
 \textsc{V.\,D. Lakhno},
 \jr{Phys. Chem. Chem. Phys} \textbf{4}, 2246 (2002).


\bibitem{CDCH}
 \textsc{A.\,W. Chin},  \textsc{A.~Datta},  \textsc{F.~Caruso},  \textsc{S.\,F.
  Huelga},  and  \textsc{M.\,B. Plenio},
 \jr{New Journal of Physics} \textbf{12}(6), 065002 (2010).


\bibitem{B0}
 \textsc{M.\,V. Berry},
 \jr{Proc. R. Soc. London A} \textbf{392}, 45 (1984).


\bibitem{B}
 \textsc{M.\,V. Berry},
 \jr{Czech. J. Phys.} \textbf{54}, 1039 (2004).


\bibitem{GABS}
 \textsc{Y.\,M. Galperin},  \textsc{B.\,L. Altshuler},  \textsc{J.~Bergli},
  \textsc{D.~Shantsev},  and  \textsc{V.~Vinokur},
 \jr{Phys. Rev. B} \textbf{76}, 064531 (2007).


\bibitem{BGA}
 \textsc{J.~Bergli},  \textsc{Y.\,M. Galperin},  and  \textsc{B.\,L.
  Altshuler},
 \jr{New Journal of Physics} \textbf{11}, 025002 (2009).


\othercit
\bibitem{abr}
 \textsc{M.~Abramowitz} and  \textsc{I.\,A. Stegun} (eds.),
{Handbook of Mathematical Functions} (Dover, New York, 1965).


\bibitem{TMT}
 \textsc{M.~Takano},  \textsc{T.~Takahashi},  and  \textsc{K.~Nagayama},
 \jr{Phys. Rev. Lett.} \textbf{80}, 5691 (1998).


\bibitem{CBC}
 \textsc{P.~Carlini},  \textsc{A.~Bizzari},  and
  \textsc{S.~Cannistrato},
 \jr{Physica D} \textbf{165}, 242 (2002).


\bibitem{JBS}
 \textsc{M.~Joyeux},  \textsc{S.~Buyukdagli},  and  \textsc{M.~Sanrey},
 \jr{Phys. Rev. E} \textbf{75}, 061914 (2007).


\bibitem{NAI}
 \textsc{A.\,I. Nesterov} and  \textsc{F.~Aceves\,de\,la Cruz},
 \jr{J. Phys. A: Math. Theor.} \textbf{41}, 485304 (2008).


\bibitem{NBB}
 \textsc{A.\,I. Nesterov},  \textsc{G.\,P. Berman},  and  \textsc{A.\,R.
  Bishop},
 \jr{IJQI} \textbf{6}, 895 (2010).


\bibitem{WW}
 \textsc{V.\,F. Weisskopf} and  \textsc{E.\,P. Wigner},
 \jr{Z. Physics} \textbf{63}, 54 (1930).


\end{thebibliography}
\end{document}